# Low-Latency Successive-Cancellation List Decoders for Polar Codes with Multi-bit Decision

Bo Yuan and Keshab K. Parhi, *Fellow*, *IEEE*

*Abstract*—**Polar codes, as the first provable capacity-achieving error-correcting codes, have received much attention in recent years. However, the decoding performance of polar codes with traditional successive-cancellation (SC) algorithm cannot match that of the low-density parity-check (LDPC) or turbo codes. Because SC list (SCL) decoding algorithm can significantly improve the error-correcting performance of polar codes, design of SCL decoders is important for polar codes to be deployed in practical applications. However, because the prior latency reduction approaches for SC decoders are not applicable for SCL decoders, these list decoders suffer from the long latency bottleneck. In this paper, we propose a multi-bit-decision approach that can significantly reduce latency of SCL decoders. First, we present a reformulated SCL algorithm that can perform intermediate decoding of 2 bits together. The proposed approach, referred as *2-bit reformulated SCL* (*2b-rSCL*) *algorithm*, can reduce the latency of SCL decoder from (3$n$-2) to (2$n$-2) clock cycles without any performance loss. Then, we extend the idea of 2-bit-decision to general case, and propose a general decoding scheme that can perform intermediate decoding of any $2^K$ bits simultaneously. This general approach, referred as *$2^K$-bit reformulated SCL ($2^K$b-rSCL) algorithm*, can reduce the overall decoding latency to as short as $n/2^{K-2}$-2 cycles. Furthermore, based on the proposed algorithms, VLSI architectures for 2b-rSCL and 4b-rSCL decoders are synthesized. Compared with a prior SCL decoder, the proposed (1024, 512) 2b-rSCL and 4b-rSCL decoders can achieve 21% and 60% reduction in latency, 1.66 times and 2.77 times increase in coded throughput with list size 2, and 2.11 times and 3.23 times increase in coded throughput with list size 4, respectively.**

## I. Introduction

Polar codes are the first provable capacity-achieving codes [1]. Due to their explicit structure and regular encoding/decoding architectures, polar codes have received much attention in recent years. To date many works have addressed theoretical analysis [1-6] [20] and hardware implementation [7-17] [21] of polar codes.

Although it has been proved that polar codes can achieve channel capacity asymptotically, the decoding performance of polar codes with the successive cancellation (SC) algorithm [1] is inferior to that of LDPC or Turbo codes. To improve the decoding performance, a successive cancellation list (SCL) decoding algorithm was presented in [4]. Simulation results show that polar codes with the use of SCL algorithm combined with simple CRC check and systematic encoding methods can outperform the same length and rate LDPC codes [4]. As a result, the SCL algorithm is believed to be the key for decoding of polar codes to be applicable in practical systems.

However, due to the inherent serial nature of successive cancellation computation, the SCL decoders suffer from long latency and low throughput problems similar to early SC decoders. Nowadays many techniques [3][7][12-15][21] have been proposed to reduce the latency of SC decoders; however, these approaches cannot be directly used to reduce the latency of the SCL decoders. As a result, to date the known VLSI designs of SCL decoder [16-17] still incur decoding latency of 3$n$-2 clock cycles.[1]

This paper presents multi-bit-decision approaches that can reduce the latency of SCL decoders. First, *2-bit reformulated SCL* (*2b-rSCL*) algorithm, which can perform intermediate decoding of 2 bits simultaneously, is presented to reduce the overall latency from (3$n$-2) cycles to (2$n$-2) cycles. Then, by generalizing the 2-bit-decision idea, we propose a general *$2^K$-bit reformulated SCL ($2^K$b-rSCL)* algorithm. By performing intermediate decoding of $2^K$ bits together, the proposed $2^K$b-rSCL decoder has latency as short as $n/2^{K-2}$-2 cycles. In order to demonstrate the advantage of the proposed approaches, VLSI architectures of 2b-rSCL and 4b-rSCL decoders are synthesized. Compared with the prior SCL decoder, the proposed (1024, 512) 2b-rSCL and 4b-rSCL decoders can achieve 21% and 60% reduction in latency, 1.66 times and 2.77 times higher in coded throughput with list size 2, and 2.11 times and 3.23 times higher in coded throughput with list size 4, respectively.

The rest of this paper is organized as follows. Section II gives a brief review of polar codes and SCL algorithm. The proposed 2b-rSCL and $2^K$b-rSCL algorithms are presented in Section III. Section IV presents the hardware architectures of the 2b-rSCL and 4b-rSCL decoders. Hardware analysis and comparisons are discussed in Section V. Section VI draws the conclusions.

 Authors are with Department of Electrical and Computer Engineering, University of Minnesota, Twin Cities, USA. Email: {yuan0103, parhi}@umn.edu.

[1] The latency calculation in [16-17] is not 3$n$-2 but varies with code rate $R$. In order to discuss the general case, in the rest of this paper we analyze the latency without specific discussion of $R$. For the latency calculation in Section V with specific code parameters, we will consider $R$ for fair comparison.

## II. Review of Polar Codes and SCL Algorithm

### A. Encoding Process of Polar Codes

Different from other block codes, an $(n, k)$ polar code is generated in two steps. First, the $k$-bit source message is extended to an $n$-bit message $\boldsymbol{x}=(u_1, u_2,\ldots u_n)$ by padding $(n\text{-}k)$ "0" bits. Notice that because the post-decoding reliability of $n$ bit positions of $\boldsymbol{u}$ can be pre-computed in [1], the $k$ most reliable positions of $\boldsymbol{u}$ are assigned $k$ information bits and other $(n\text{-}k)$ least reliable positions are forced to be "0". Then, the $n$-bit message $\boldsymbol{u}$ is multiplied with an $n\times n$ generator matrix $\boldsymbol{G}$ to generate the transmitted codeword $\boldsymbol{x}=(x_1, x_2,\ldots,x_n)$. Fig. 1 shows the implementation of a polar code encoder with $n$=4.

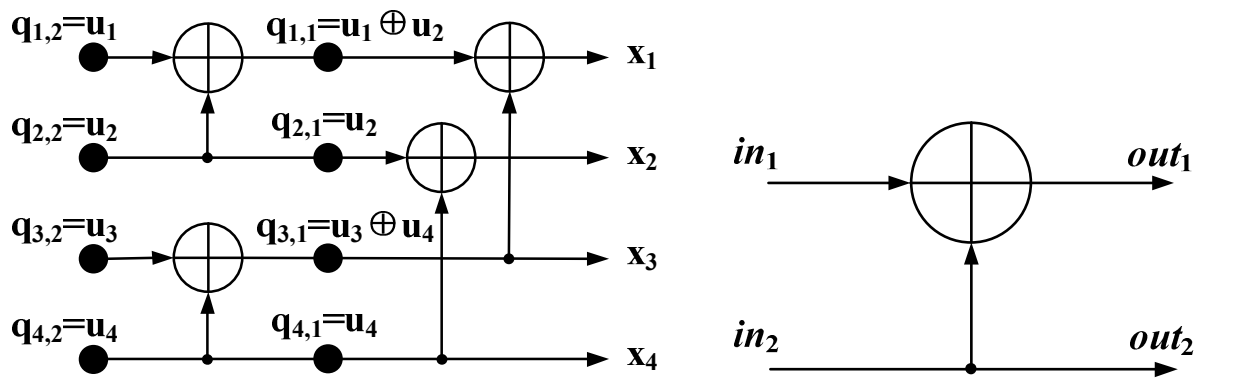

(a) (b)

Fig. 1. (a) Implementation of $n$=4 polar encoder. (b) Basic unit of polar encoder.

### B. Successive Cancellation Decoding Algorithm

At the receiver end, due to the corruption from transmission noise, the transmitted codeword $\boldsymbol{x}$ changes to the received codeword $\boldsymbol{y}=(y_1, y_2,\ldots,y_n)$. Since the required information bits are contained in $\boldsymbol{u}$, a polar code decoder is needed to recover the $\boldsymbol{u}$ from the $\boldsymbol{y}$. In [1], Arıkan proposed a successive cancellation (SC) decoder to perform this recovery. Fig. 2 shows the example decoding procedure of this SC decoder for $n$=4 polar code based on likelihood form. As seen in this figure, the SC decoder consists of $m=\log_2 n=2$ stages, where each stage consists of two types of 4-input-2-output units, referred as **f** unit and **g** unit, respectively. In addition, a 2-input-1-output hard-decision unit denoted as **h** is used at the last stage of SC decoder (stage-2) to determine the estimate of $u_i$, referred as $\hat{u}_i$. Besides, each **f** or **g** unit is labeled a number to indicate the clock cycle index when it is activated. This labeling system reveals the inherent serial nature of the SC decoding algorithm. For example, in Fig. 2 the decoded bits are output at cycles 2, 3, 5, 6, respectively.

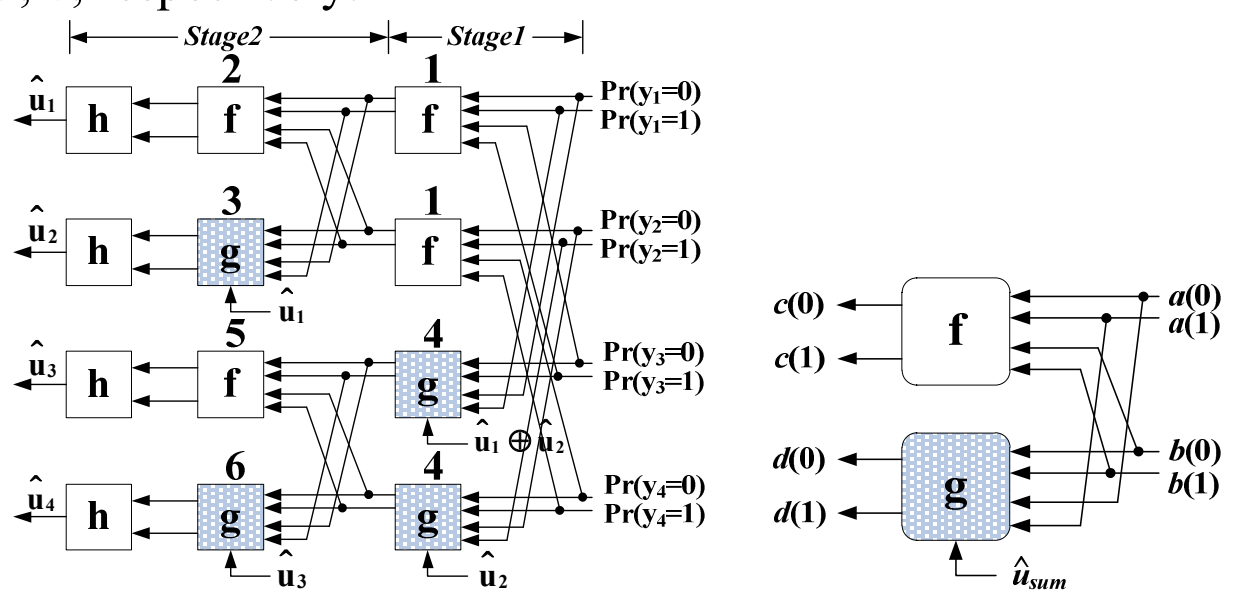

(a) (b)

Fig. 2. (a) SC decoding procedure with $n$=4. (b) Basic unit of SC decoder.

In addition, the functions of **f** and **g** units can be derived via the analogy between polar code encoder and decoder. Fig. 1 (b) and Fig. 2(b) show the general basic unit in polar encoder and decoder, respectively. For the basic unit of encoder (see Fig. 1(b)), it performs a left-to-right transformation from $in_1$ and $in_2$ to $out_1$ and $out_2$. Hence, the transformation equations in Fig. 1(b) are:

$$out_1=in_1 \oplus in_2, \text{ and } out_2=in_2, \tag{1}$$

where $\oplus$ represents the exclusive-or operation.

On the other hand, for the basic unit of decoder in Fig. 2(b), as indicated in [1], its function is just a right-to-left estimation from the likelihoods of $out_1$ and $out_2$ to the likelihoods of $in_1$ and $in_2$. Therefore, according to the left-to-right transformation equations (1), the "expected" relationship from the estimates of $out_1$ and $out_2$ to the estimates of $in_1$ and $in_2$ can be derived as:

$$\widehat{in}_1=\widehat{out}_1 \oplus \widehat{out}_2, \text{ and } \widehat{in}_2=\widehat{out}_2. \tag{2}$$

With the help of the above "guideline" equations (2), we can now develop the functions of **f** and **g** units in Fig. 2(b). First we assume the previously decoded bits $\hat{u}_1, \hat{u}_2 \ldots \hat{u}_{2i-2}$ have been determined as binary values $z_1, z_2, \ldots z_{i-1}$, respectively. For simplicity, this event is denoted as $\hat{u}_1^{i-1}=z_1^{i-1}$. Then, the two outputs of **f** unit, referred as $c(0)$ and $c(1)$, can be derived:

$$\begin{aligned} c(0) &= \Pr(\widehat{in}_1=0, \hat{u}_1^{i-1}=z_1^{i-1}) \\ &= \Pr(\widehat{out}_1=0, \hat{u}_1^{i-1}=z_1^{i-1})\Pr(\widehat{out}_2=0, \hat{u}_1^{i-1}=z_1^{i-1}) \\ &\quad + \Pr(\widehat{out}_1=1, \hat{u}_1^{i-1}=z_1^{i-1})\Pr(\widehat{out}_2=1, \hat{u}_1^{i-1}=z_1^{i-1}) \\ &= a(0)b(0)+a(1)b(1) \end{aligned} \tag{3}$$

$$\begin{aligned} c(1) &= \Pr(\widehat{in}_1=1, \hat{u}_1^{i-1}=z_1^{i-1}) \\ &= \Pr(\widehat{out}_1=0, \hat{u}_1^{i-1}=z_1^{i-1})\Pr(\widehat{out}_2=1, \hat{u}_1^{i-1}=z_1^{i-1}) \\ &\quad + \Pr(\widehat{out}_1=1, \hat{u}_1^{i-1}=z_1^{i-1})\Pr(\widehat{out}_2=0, \hat{u}_1^{i-1}=z_1^{i-1}) \\ &= a(0)b(1)+a(1)b(0) \end{aligned} \tag{4}$$

where $a(0)$, $a(1)$, $b(0)$, $b(1)$ are the inputs of **f** or **g** unit..

Due to the successive property of SC algorithm, $d(0)$ and $d(1)$, as the outputs of **g** unit, are determined by the estimate of $in_1$. When it is 0, according to (2), we have:

$$d(0)=\Pr(\widehat{in}_2=0, \widehat{in}_1=0, \hat{u}_1^{i-1}=z_1^{i-1})$$

$$=\Pr(\widehat{out}_1=0, \hat{u}_1^{i-1}=z_1^{i-1})\Pr(\widehat{out}_2=0, \hat{u}_1^{i-1}=z_1^{i-1})=a(0)b(0) \tag{5}$$

$$d(1)=\Pr(\widehat{in}_2=1, \widehat{in}_1=0, \hat{u}_1^{i-1}=z_1^{i-1})$$

$$=\Pr(\widehat{out}_1=1, \hat{u}_1^{i-1}=z_1^{i-1})\Pr(\widehat{out}_2=1, \hat{u}_1^{i-1}=z_1^{i-1})=a(1)b(1) \tag{6}$$

Similarly, when $\widehat{in}_1$ is 1, according to (2), we have:

$$d(0)=\Pr(\widehat{in}_2=0, \widehat{in}_1=1, \hat{u}_1^{i-1}=z_1^{i-1})$$

$$=\Pr(\widehat{out}_1=1, \hat{u}_1^{i-1}=z_1^{i-1})\Pr(\widehat{out}_2=0, \hat{u}_1^{i-1}=z_1^{i-1})=a(1)b(0) \tag{7}$$

$$d(1)=\Pr(\widehat{in}_2=1, \widehat{in}_1=1, \hat{u}_1^{i-1}=z_1^{i-1})$$

$$=\Pr(\widehat{out}_1=0, \hat{u}_1^{i-1}=z_1^{i-1})\Pr(\widehat{out}_2=1, \hat{u}_1^{i-1}=z_1^{i-1})=a(0)b(1) \tag{8}$$

As a result, by summarizing (5)(6)(7)(8), we can obtain the unified function for **g** unit:

$$d(0)=\Pr(\widehat{in}_2=0, \widehat{in}_1=\hat{u}_{sum}, \hat{u}_1^{i-1}=z_1^{i-1})=a(\hat{u}_{sum})b(0) \tag{9}$$

$$d(1)=\Pr(\widehat{in}_2=1, \widehat{in}_1=\hat{u}_{sum}, \hat{u}_1^{i-1}=z_1^{i-1})=a(1-\hat{u}_{sum})b(1) \tag{10}$$

Besides, for **h** unit, since it is the hard-decision unit, we can obtain its function as follows:

$$\hat{u}_i = \begin{cases} 0 & \text{if } a(0) \geq a(1) \text{ or } \hat{u}_i \text{ is frozen bit} \\ 1 & \text{if } a(0) < a(1) \text{ and } \hat{u}_i \text{ is free bit} \end{cases} \tag{11}$$

In general, equations (3)(4)(9)(10)(11) describe the likelihood-based SC algorithm.

On the other hand, from the view of code tree, the SC algorithm can be described as a path searching process. Fig. 3 shows an example for $n$=4 and $k$=4 SC decoding procedure over the code tree. This $n$=4 code tree consists of 4 levels, where each level represents a decoded bit. The value associated with each node is the likelihood-based metric for the decoding path from root node to the current node. For example, 0.33 on the leftmost side indicates that for the path $\hat{u}_1$=0 and $\hat{u}_2$=0, denoted as the length-2 path (00), its metric is given by Pr($\hat{u}_1$=0, $\hat{u}_2$=0)=0.33. For the 0.12 on the rightmost side, it indicates the metric for path $\hat{u}_1$=1, $\hat{u}_2$=1 and $\hat{u}_3$=1, denoted as the length-3 path (111), is given by Pr($\hat{u}_1$=1, $\hat{u}_2$=1, $\hat{u}_3$=1)=0.12. In particular, the path metrics associated with the nodes at the lowest level (level 4) represent the different likelihoods for the different combinations of ($\hat{u}_1\,\hat{u}_2\,\hat{u}_3\,\hat{u}_4$). The valid output of this $n$=4 SC polar decoder should be the length-4 path which has the largest metric at the lowest level. In this example it is (0010) with path metric Pr($\hat{u}_1$=0, $\hat{u}_2$=0, $\hat{u}_3$=1 $\hat{u}_4$=0)=0.19.

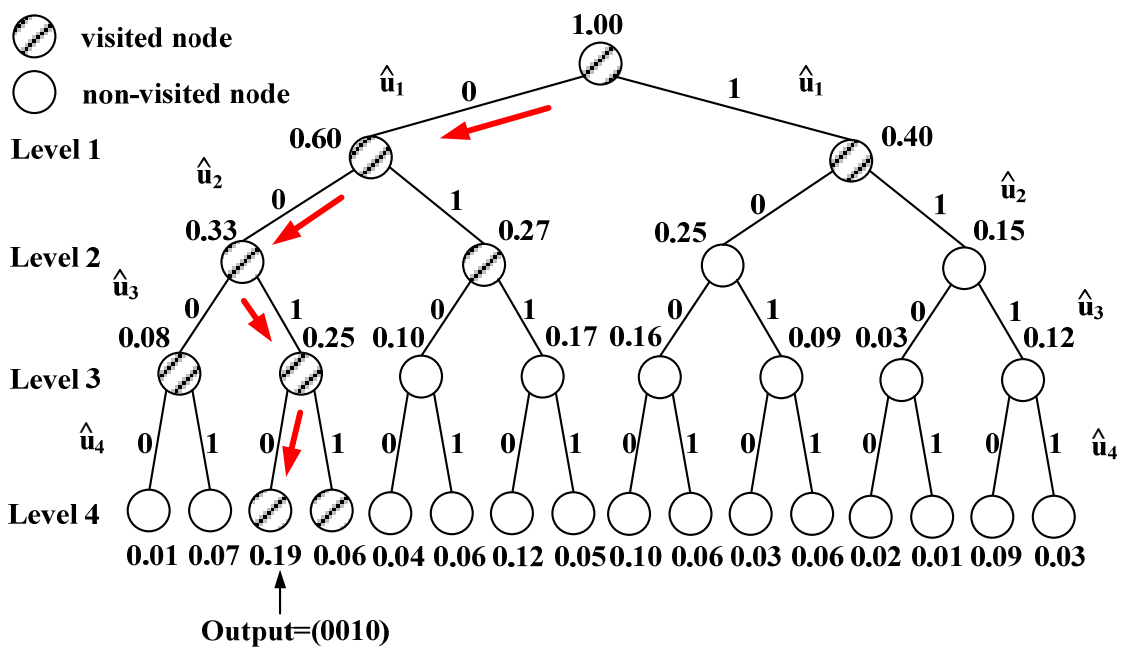

Fig. 3. Searching process of SC decoder over code tree with $n$=4 and $k$=4.

Notice the aforementioned path metrics are calculated by the **f** or **g** units in the last stage of the SC decoder (for example stage-2 in Fig. 2(a)). For the length-$i$ path, its path metric is computed by the **f** or **g** unit associated with $\hat{u}_1$. For example, for $n$=4 polar code, the path metric for path ($\hat{u}_1\,\hat{u}_2$) is computed by index-3 **g** unit in Fig. 2(a). Similarly, the path metric for path ($\hat{u}_1\,\hat{u}_2\,\hat{u}_3$) is computed by index-5 **f** unit in Fig. 2(a).

In order to find the decoding path with the largest metric, SC algorithm adopts a locally optimal searching strategy. As shown in Fig. 3, the arrows represent the survival decoding path of the SC decoder. In the $i$-th level, the SC decoder first visits the two children nodes (striped nodes in Fig. 3) that are connected to the current survival length-($i$-1) path. Since the metrics of length-$i$ paths are associated with these children nodes, the SC decoder then can obtain the metrics of length-$i$ paths. After comparing the metrics, the SC decoder only selects the length-$i$ path which has the larger metric as the updated survival path, while the path which has the smaller metric will never be explored in the future. Based on this searching strategy, in Fig. 3 the length-4 path (0010) with metric 0.19 is selected as the output of SC decoder. In this example, the SC decoder works well since it finds the valid length-4 path with the largest metric.

### C. *Successive Cancellation List Decoding Algorithm*

An essential drawback of the SC algorithm is that its searching strategy over the code tree is only locally optimal, but not globally optimal. As a result, in many cases the ($n$, $k$) SC decoder cannot find the length-$n$ path with the largest metric. For example, if we apply SC decoding approach in Fig. 4, its output is (0010) with metric 0.19; however, the valid length-4 path with largest metric should be (1000) with metric 0.23.

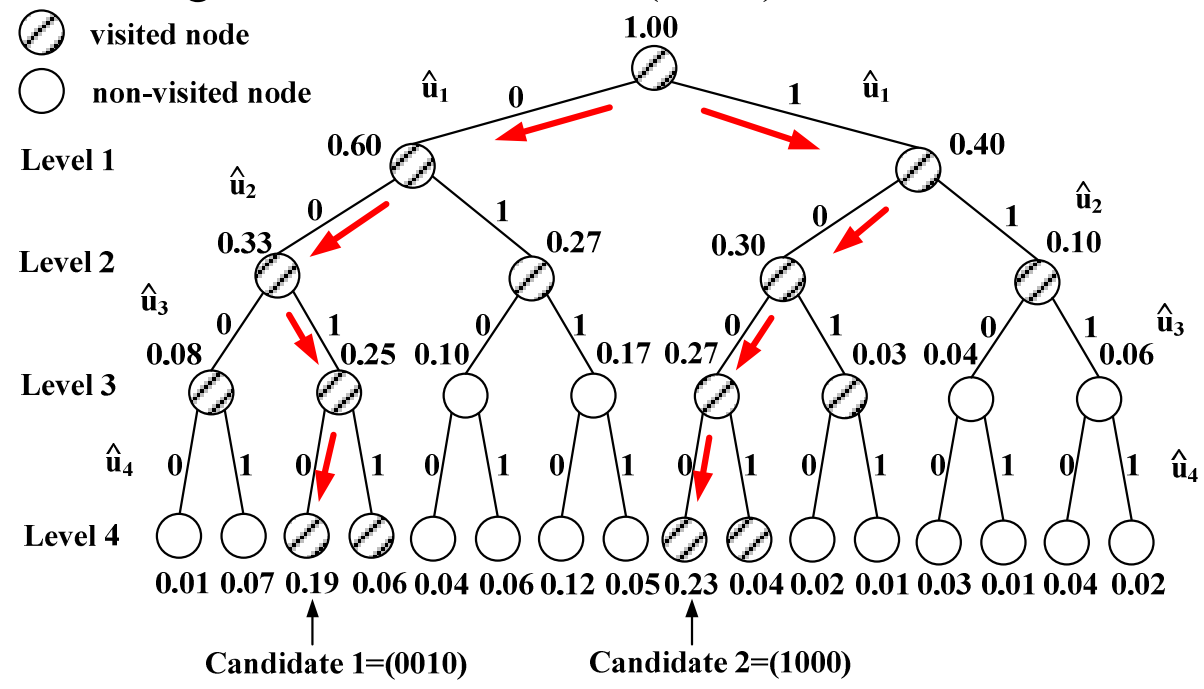

Fig. 4. Searching process of SCL decoder with $n$=4, $k$=4 and $L$=2.

The reason for the inefficiency of SC algorithm in this example is that sometimes the unexplored path, instead of the chosen survival path, has the larger path metric. Based on this observation, successive cancellation list (SCL) algorithm [4] was proposed to perform searching process along multiple survival paths at the same time. Here the maximum number of the survival paths is referred as the list size ($L$). Fig. 4 shows an example for the $n$=4 and $k$=4 SCL decoder with $L$=2. As shown in Fig. 4, at the $i$-th level, the SCL decoder visits all the 2$L$ children nodes (striped nodes in Fig. 4) that are connected to the length-($i$-1) survival paths. After calculating all the 2$L$ new path metrics associated with these children nodes, the SCL decoder selects the $L$ length-$i$ paths which have the larger metrics as the updated survival paths. From Fig. 4 it can be seen that the valid decoding path (1000), which could not be traced by SC decoder before, now can be found by the SCL decoder.

## III. The Proposed Reformulated SCL Algorithms

### A. *Long latency problem of original SCL decoder*

In general, the SCL algorithm can improve decoding performance significantly over the SC algorithm [4]. However, one of the major challenges for the practical use of SCL decoder is the long latency problem. Because an $L$-size ($n$, $k$) SCL decoder can be viewed as the combination of $L$ copies of ($n$, $k$) SC component decoders (see Fig. 5), an ($n$, $k$) SCL decoder needs the same (2$n$-2) cycles to process its **f** and **g** units as its SC component decoders do. In addition, since SCL decoders need to sort 2$L$ path metrics and select $L$ largest metrics for each decoded bit (see Fig. 5), extra $n$ cycles are needed to carry out the sorting and selecting function to avoid long critical path [16]. Therefore, the latency of an ($n$, $k$) SCL decoder is 3$n$-2 cycles. As discussed in Section I, although some methods have been proposed to reduce the latency of SC decoders, these approaches cannot be directly applied to the SCL decoder. As a result, the latency of current known SCL decoder [16-17] is still very long. Table I shows an example decoding scheme of conventional SCL decoder for $n$=4 polar code. Here in this table the symbols **f** and **g** represent the **f** and **g** units in each SC component decoder of Fig. 2, respectively. Besides, the symbol **s** represents the path metrics sorting and selecting operation for each intermediately decoded bit.

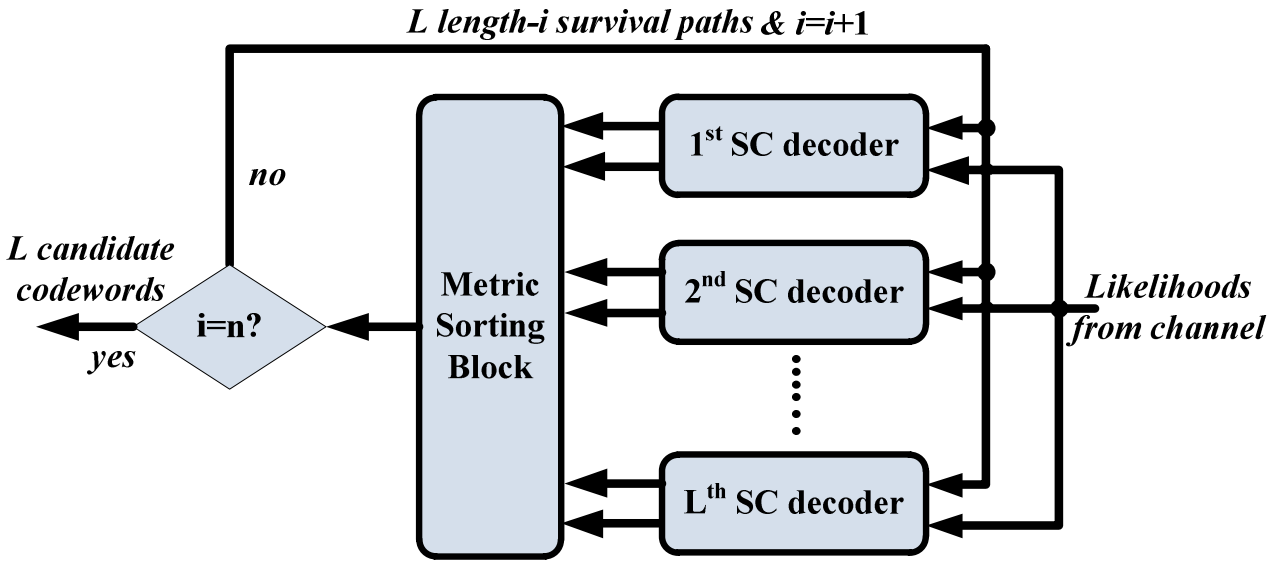

Fig. 5. Block diagram of $L$-size SCL decoder.

TABLE I. DECODING SCHEME OF CONVENTIONAL SCL DECODER WITH N=4

| Clock cycle | 1 | 2 | 3 | 4 | 5 | 6 | 7 | 8 | 9 | 10 |
|---|---|---|---|---|---|---|---|---|---|---|
| Stage-1 | f | | | | | g | | | | |
| Stage-2 | | f | s | g | s | | f | s | g | s |
| Bit decision | | | $\hat{u}_1$ | | $\hat{u}_2$ | | | $\hat{u}_3$ | | $\hat{u}_4$ |

*B. 2-bit reformulated SC List (2b-rSCL) Algorithm*

As seen in Table I, more than 60% latency of SCL decoder is due to the computation of **f**, **g** and **s** in the last stage (stage-2, in Table I). This phenomenon implies that the reduction of latency in the last stage can lead to significant reduction of the overall latency of SCL decoder. Therefore, in this sub-section we propose to reformulate the original computation of the last stage. This reformulated computation in the last stage can save many clock cycles without any performance loss.

Table I shows that the computation of the last stage can be viewed as multiple "**f s g s**" functions to perform intermediate decoding of two consecutive bits $\hat{u}_{2i-1}$ and $\hat{u}_{2i}$. Since the **f**/**g** units and **s** in the last stage contribute to path metrics calculation and selection, respectively, hence the goal of our reformulation on the last stage is to find a simplified method that can compute path metrics and sort/select them to perform intermediate decoding of $\hat{u}_{2i-1}$ and $\hat{u}_{2i}$ more quickly.

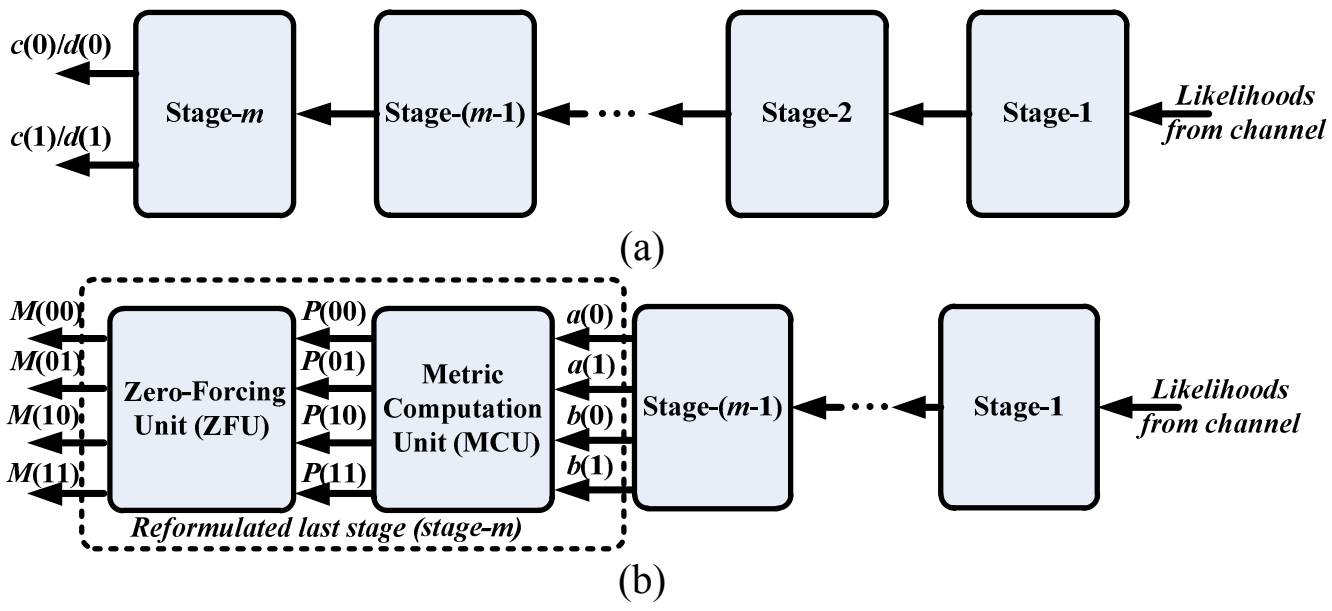

Fig. 6. Block diagram of (a) original SC component decoder of SCL decoder. (b) reformulated SC component decoder of 2b-rSCL decoder.

Fig. 6 (a) and (b) show the block diagram of the original and reformulated SC component decoder for SCL decoding, respectively. From these two figures it can be seen that the reformulated SC decoder replaces the original stage-$m$ with two new units, referred as metric computation unit (MCU) and zero-forcing unit (ZFU), respectively. Besides that, as shown in Fig. 5, a sorting block (**s** symbol in Table I) is also needed to sort the path metrics output from all the $L$ SC component decoders. Because the sorting block is an individual block that does not belong to any SC component decoder, in this subsection we do not discuss sorting block but focus on the functions of MCU and ZFU. The architecture of sorting block will be presented in Section IV.

*Metric Computation Unit (MCU)*

As shown in Fig. 7, metric computation unit (MCU) calculates the likelihoods for different combinations of $\hat{u}_{2i-1}$ and $\hat{u}_{2i}$ with the use of the messages $a(0)$, $a(1)$, $b(0)$ and $b(1)$ output from stage-($m$-1). The principle of this calculation can be derived from (5)-(8). Since for the last stage of each SC component decoder, $\hat{u}_{2i-1}$ and $\hat{u}_{2i-1}$ are the estimates of $in_1$ and $in_2$, respectively, therefore by making $\hat{in}_1 = \hat{u}_{2i-1}$ and $\hat{in}_2 = \hat{u}_{2i}$ in (5)-(8) we can have:

$$P(00) \triangleq \Pr(\hat{u}_{2i-1}=0,\ \hat{u}_{2i}=0, \hat{u}_1^{2i-2}=z_1^{2i-2}) = a(0)b(0)$$
$$P(01) \triangleq \Pr(\hat{u}_{2i-1}=0,\ \hat{u}_{2i}=1, \hat{u}_1^{2i-2}=z_1^{2i-2}) = a(1)b(1)$$
$$P(10) \triangleq \Pr(\hat{u}_{2i-1}=1,\ \hat{u}_{2i}=0, \hat{u}_1^{2i-2}=z_1^{2i-2}) = a(1)b(0)$$
$$P(11) \triangleq \Pr(\hat{u}_{2i-1}=1,\ \hat{u}_{2i}=1, \hat{u}_1^{2i-2}=z_1^{2i-2}) = a(0)b(1) \quad (12)$$

where $\hat{u}_1^{2i-2}=z_1^{2i-2}$ denotes that the previously decoded bits $\hat{u}_1$, $\hat{u}_2 \ldots \hat{u}_{2i-2}$ are assumed to have been determined as $z_1, z_2, \ldots z_{2i-2}$, respectively.

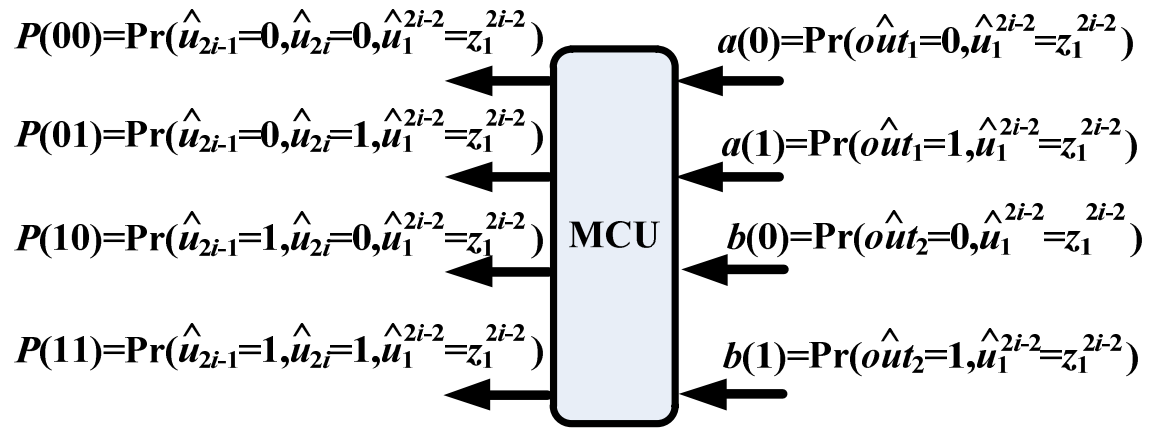

Fig. 7. Block diagram of MCU for 2b-rSCL decoder.

Equations (12) describe the calculation of the joint likelihoods of $\hat{u}_{2i-1}$, $\hat{u}_{2i}$ and $\hat{u}_1^{2i-2}=z_1^{2i-2}$. Now we show that these joint likelihoods are just the actual metrics of length-$2i$ paths. Consider one of the current length-($2i$-2) survival path in the code tree as ($\hat{u}_1 \ldots \hat{u}_{2i-2}$)=($z_1 \ldots z_{2i-2}$). As shown in Fig. 8, with the different combination of $\hat{u}_{2i-1}$ and $\hat{u}_{2i}$, this length-($2i$-2) path can be extended to four length-($2i$) paths as ($\hat{u}_1 \ldots \hat{u}_{2i-2}\ \hat{u}_{2i-1}\ \hat{u}_{2i}$)=($z_1 \ldots z_{2i-2}pq$), where $p$ and $q$ are binary 0 or 1. According to the definition of path metric, with the four combinations of $p$ and $q$, Pr($\hat{u}_{2i-2}$=$p$, $\hat{u}_{2i-2}$=$q$, $\hat{u}_1^{2i-2}=z_1^{2i-2}$) in (12) are just the actual metrics of the above four extended length-($2i$) paths. As a result, according to equations (12), with the knowledge of $a(0)$, $a(1)$, $b(0)$ and $b(1)$ output from the stage-($m$-1), we can directly obtain the actual path metrics of four length-($2i$) paths.($\hat{u}_1 \ldots \hat{u}_{2i-2}\ \hat{u}_{2i-1}\ \hat{u}_{2i}$)=($z_1 \ldots z_{2i-2}pq$).

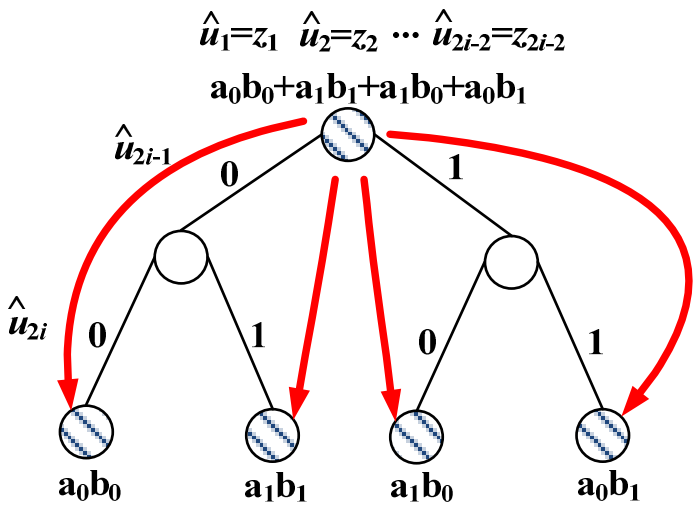

Fig. 8. Extension from one length-($2i$-2) path to four length-($2i$) paths with direct computation of actual path metrics.

*Zero-Forcing Unit (ZFU)*

Although equations (12) provide a fast approach to compute the actual metrics of length-$2i$ paths, a post-processing operation is still needed before inputting those calculated metrics into the sorting block. This is because the values of $\hat{u}_{2i-1}$ and $\hat{u}_{2i}$ do not only depend on the corresponding path metrics, but also on whether they are frozen bits or not. Notice that when the current decoded bit $\hat{u}_i$ is a frozen bit, the paths with $\hat{u}_i$=1 are not qualified and should never be selected even if they have larger metrics than their counterparts. As a result, in order to avoid selecting those unqualified paths, we need a zero-forcing unit (ZFU) to force the metrics of those unqualified paths to 0. The reason of this zero-forcing operation is that since the SCL decoder only selects the $L$ survival paths with larger metrics for each $\hat{u}_i$, therefore the unqualified paths with metric values 0 will never be classified into the group of $L$ paths with larger metrics. As a result, the validity of the function is guaranteed.

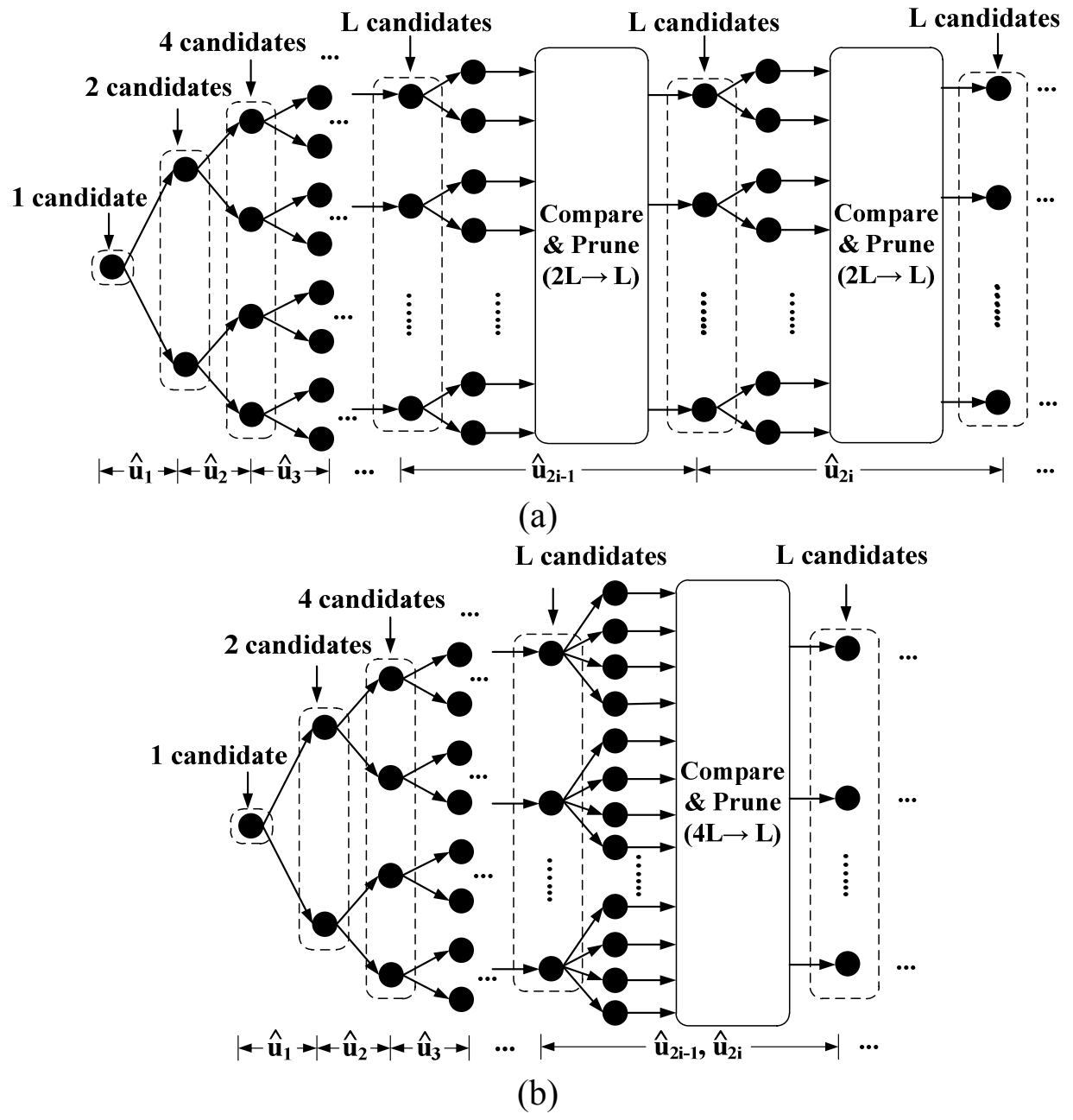

Fig. 9. $L$-size decoding scheme of (a) original SCL decoder. (b)2b- rSCL decoder.

Since the proposed reformulated last stage involves both $\hat{u}_{2i-1}$ and $\hat{u}_{2i}$, the function of ZFU is derived as follows:

Assign $M(pq)$=$P(pq)$ for path $(z_1^{2i-2}pq)$ with $p$, $q \in \{0,1\}$;

If $\hat{u}_{2i\text{-}1}$ is frozen, then reassign $M(10)$=0 and $M(11)$=0;

If $\hat{u}_{2i}$ is frozen, then reassign $M(01)$=0 and $M(11)$=0. (13)

Equations (12) and (13) describe the reformulated function of the last stage of SC component decoder. With the help of this reformulation, $\hat{u}_{2i-1}$ and $\hat{u}_{2i}$, as the two successive decoded bits, can now be intermediately decoded at the same time. Fig. 9 (a) and (b) show the decoding procedure of original SCL decoding and the proposed reformulated approach with list size $L$, respectively. In the conventional SCL algorithm, with the comparison of their metrics, the $L$ length-($2i$-1) survival paths are selected from $2L$ candidates for each time. And each selection can only perform intermediate decoding of one decoded bit (see Fig. 9(a)). Instead, in the reformulated approach, the $L$ length-($2i$) survival paths are selected from $4L$ candidates for each time. As a result, the two successive bits can now be intermediately decoded simultaneously in each selection (see Fig. 9(b)).

Considering the proposed reformulation can allow two bits to be intermediately decoded at the same time, this new SCL algorithm is referred as *2-bit reformulated SCL* (*2b-rSCL*) algorithm and described in Scheme-A.

**Scheme A: 2-bit SCL decoding (2b-rSCL) with list size L for (n, k) polar codes**

**Input:** *Likelihoods of each bit in the received codeword*
**For** $i=1$ **to** $n/2$
  **For each length-**$(2i-2)$ **survival path** $(\hat{u}_1 \ldots \hat{u}_{2i-2}) \triangleq z_1^{2i-2}$
    **SC decoding:**
      *Activate stage* $-1 \sim$ *stage* $-(m-1)$ *of SC component decoder*
      *stage* $-(m-1)$ *outputs* $a(0), a(1), b(0), b(1)$
    **Path Expansion:**
      *Expand survival path* $z_1^{2i-2}$ *to 4 candidate paths* $(z_1^{2i-2}\hat{u}_{2i-1}\hat{u}_{2i})$:
      *1 length - (2i - 2) path* $\Rightarrow$ *4 length - (2i) paths*
    **Metric Computation:**
      *Calculate actual path metrics* $P(pq)$ *for* 4 *length* $-(2i)$ *paths*:
        $P(00)$=$a(0)b(0)$ *for path* $(z_1^{2i-2}00)$, $P(01)$=$a(1)b(1)$ *for path* $(z_1^{2i-2}01)$,
        $P(10)$=$a(1)b(0)$ *for path* $(z_1^{2i-2}10)$, $P(11)$=$a(0)b(1)$ *for path* $(z_1^{2i-2}11)$
    **Forcing Zero:**
      *Calculate the new path metrics* $M(pq)$ *with forcing - zero operation*:
        $M(pq) = P(pq)$ *for path* $(z_1^{2i-2}pq)$ *with* $p, q \in \{0,1\}$
        *if* $\hat{u}_{2i\text{-}1}$ *is frozen, then* $M(10)=0$ *and* $M(11)=0$
        *if* $\hat{u}_{2i}$ *is frozen, then* $M(01)=0$ *and* $M(11)=0$
  **End for**
  **Compare and Prune:**
    *Compare the metrics* $M(pq)$ *of all the 4L length - (2i) candidate paths*
    *Select L paths with the L largest metrics as the new survival paths*
**End for**
**Output:** *Choose the length - n survival path with the largest metric*

The proposed 2b-rSCL algorithm can greatly reduce the latency of the original SCL decoder. Recall that in the original searching procedure (see Fig. 4), the SCL decoder needs to compute the path metrics associated with the striped nodes in each level of the code tree. On the other hand, since the 2b-rSCL only needs to compute the metrics for length-($2i$) paths, the metrics computation for length-($2i$-1) paths are totally avoided (see Fig. 8). As a result, for the same code tree the 2b-rSCL decoder only needs to visit the striped nodes at even levels instead of at all the levels. For example, by comparing Fig. 4 and Fig. 10, it can be found that the reformulated SCL decoder does not need to visit the nodes at level 1 and level 3 anymore. As a result, this new decoding scheme leads to immediate saving in clock cycles.

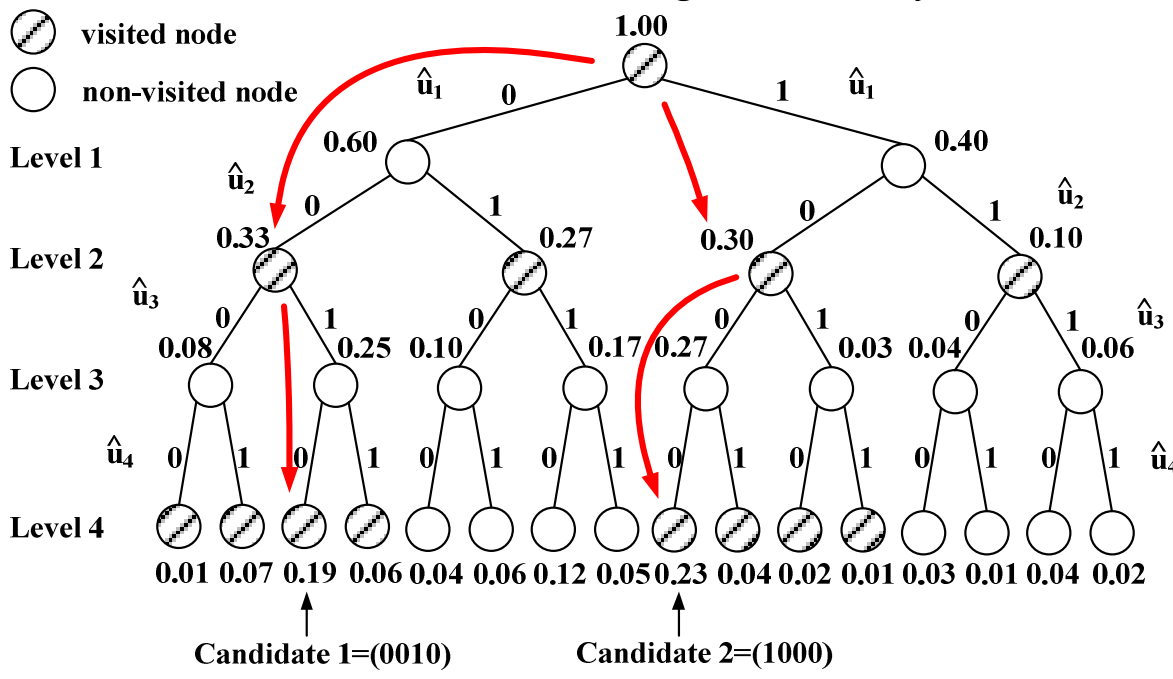

Fig. 10. Searching process of 2b-rSCL decoder over code tree with $n$=4, $k$=4 and $L$=2.

Table II shows the example decoding scheme of the proposed 2b-rSCL decoder with $n$=4. Here **mc&zf** in Table II denotes the *metric computation* and *zero-forcing* operations, which are described from line 10 to line 18 of Scheme-A in detail. Compared with the scheme of conventional SCL decoder (see Table I), it can be seen that the reformulation at the last stage (stage-2 in this example) leads to significant reduction in clock cycles. For the intermediate decoding of each two successive bits $\hat{u}_{2i-1}$ and $\hat{u}_{2i}$, the original SCL decoder (see Table I) needs 4 cycles (**f**, **s**, **g**, **s**), while the 2b-rSCL decoder in Table II only needs 2 cycles (**mc&zf**, **s**). In general, for an ($n$, $k$) polar code, the overall latency of 2b-rSCL decoder can reduce from 3$n$-2 to 2$n$-2 clock cycles.

TABLE II. DECODING SCHEME OF 2B-SCL DECODER WITH N=4

| Clock Cycle | 1 | 2 | 3 | 4 | 5 | 6 |
|---|---|---|---|---|---|---|
| Stage-1 | f | | | g | | |
| Stage-2 | | mc&zf | s | | mc&zf | s |
| Bit decision | | | $\hat{u}_1$ $\hat{u}_2$ | | | $\hat{u}_3$ $\hat{u}_4$ |

*C. $2^K$-bit reformulated SC List ($2^K$b-rSCL) Algorithm*

In subsections III-B we presented 2b-rSCL algorithm that can perform intermediate decoding of 2 bits at the same time. In this subsection, we extend the prior approach to a more general case, and propose a new algorithm, referred as *$2^K$-bit reformulated SC List ($2^K$b-rSCL)*, which can perform intermediate decoding of $2^K$ bits simultaneously.

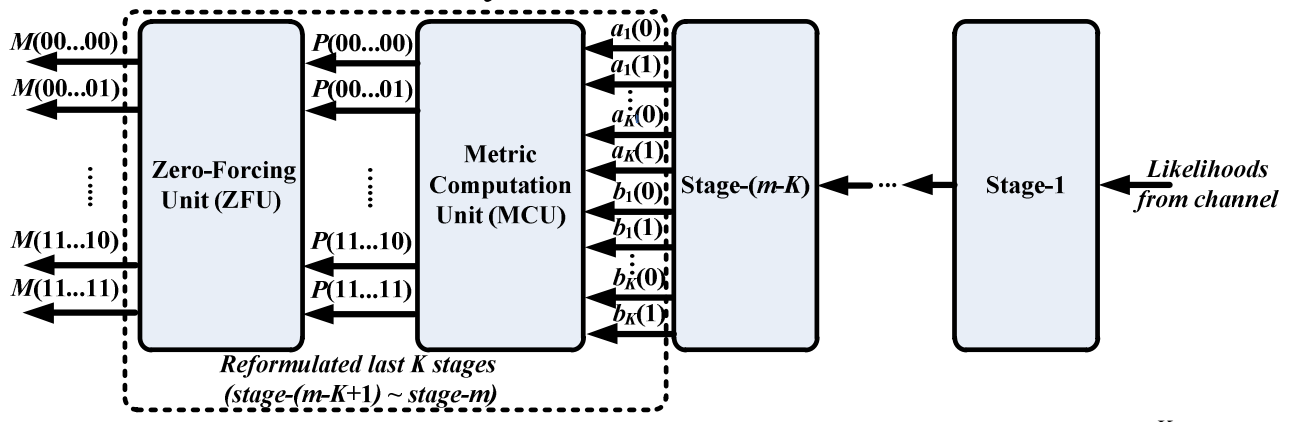

Fig. 11. Block diagram of reformulated SC component decoder of $2^K$b-rSCL decoder.

As shown in Fig. 11, the $2^K$b-rSCL decoder reformulates the last $K$ stages of original SCL decoder. Similar to the case in 2b-rSCL decoder, the reformulated part of $2^K$b-rSCL decoder consists of MCU and ZFU as well.

*Metric Computation Unit (MCU)*

The function of MCU in $2^K$b-rSCL decoder is to compute the joint probabilities of $2^K$ successive bits as $\hat{u}_{2^K(i-1)+1}$, $\hat{u}_{2^K(i-1)+2}$, … and $\hat{u}_{2^K i}$. Similar to the discussion in 2-bit-decision case, we first investigate the transformation of $u_{2^K(i-1)+1} \dots u_{2^K i}$.

As shown in Fig. 12, the transformation of $2^K$ successive bits can be viewed as the multiplication with matrix $\boldsymbol{U}$, where $\boldsymbol{U}$ is $2^K \times 2^K$ generator matrix $\boldsymbol{G}$.

Denote $\vec{\boldsymbol{u}}_{2^K,i} \triangleq (u_{2^K(i-1)+1}, u_{2^K(i-1)+2}, \dots, u_{2^K i})$ and $\overrightarrow{\boldsymbol{out}}_{2^K} \triangleq (out_1, out_2, \dots out_{2^K})$, then we have:

$$\overrightarrow{\boldsymbol{out}}_{2^K} = \vec{\boldsymbol{u}}_{2^K,i}\boldsymbol{U} \tag{14}$$

In particular, if we denote the $j$-th column vector of $\boldsymbol{U}$ as $\boldsymbol{U(j)}$, then according to (23) we have:

$$out_j = \vec{\boldsymbol{u}}_{2^K,i}\,\boldsymbol{U(j)} \tag{15}$$

Equation (24) describes the left-to-right transformation of the $u_{2^K(i-1)+1}, u_{2^K(i-1)+2}, \dots$ and $u_{2^K i}$ in encoding phase.

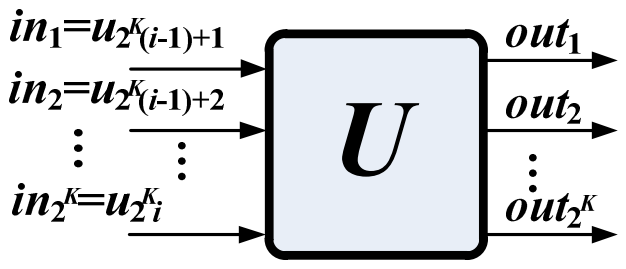

Fig. 12. Encoding procedure for $u_{2^K(i-1)+1}, u_{2^K(i-1)+2}, \dots$ and $u_{2^K i}$.

Then, based on (15), the right-to-left "guideline" in decoding procedure should be:

$$\hat{\vec{\boldsymbol{u}}}_{2^K,i} = \widehat{\overrightarrow{\boldsymbol{out}}}_{2^K}\boldsymbol{U}^{-1} = \widehat{\overrightarrow{\boldsymbol{out}}}_{2^K}\boldsymbol{U} \tag{16}$$

where $\hat{\vec{\boldsymbol{u}}}_{2^K,i} \triangleq (\hat{u}_{2^K(i-1)+1}, \hat{u}_{2^K(i-1)+2}, \dots, \hat{u}_{2^K i})$ and $\widehat{\overrightarrow{\boldsymbol{out}}}_{2^K} \triangleq (\widehat{out}_1, \widehat{out}_2, \dots \widehat{out}_{2^K})$, respectively.

According to (15) and (16), we have:

$$\widehat{out}_j = \hat{\vec{\boldsymbol{u}}}_{2^K,i}\,\boldsymbol{U(j)} \text{ and } \hat{u}_{2^K(i-1)+1} = \widehat{\overrightarrow{\boldsymbol{out}}}_{2^K}\boldsymbol{U(j)}. \tag{17}$$

Note that in (16) we use the special property that $\boldsymbol{U}^{-1}=\boldsymbol{U}$.

As shown in Fig. 13, the inputs of MCU are $a_1(0)$, $a_1(1)$,…, $a_{2^{K-1}}(0)$, $a_{2^{K-1}}(1)$, $b_1(0)$, $b_1(1)$,…, $b_{2^{K-1}}(0)$ and $b_{2^{K-1}}(1)$, respectively. With the help of (17), we can obtain the joint probabilities of $\hat{u}_{2^K(i-1)+1}$, $\hat{u}_{2^K(i-1)+2}$, … and $\hat{u}_{2^K i}$ as follows.

$$
\begin{aligned}
&P(\alpha_1\alpha_2\alpha_3\dots\alpha_{2^K})\\
&\triangleq \Pr(\hat{u}_{2^K(i-1)+1}=\alpha_1, \hat{u}_{2^K(i-1)+2}=\alpha_2, \dots, \hat{u}_{2^K i}=\alpha_{2^K}, \hat{u}_1^{2^K(i-1)}=z_1^{2^K(i-1)})\\
&= \Pr(\widehat{out}_1 = \hat{\vec{\boldsymbol{u}}}_{2^K,i}\boldsymbol{U}(\boldsymbol{1}), \widehat{out}_2 = \hat{\vec{\boldsymbol{u}}}_{2^K,i}\boldsymbol{U}(\boldsymbol{2}), \dots, \widehat{out}_{2^K} = \hat{\vec{\boldsymbol{u}}}_{2^K,i}\boldsymbol{U}(\boldsymbol{2^K}),\\
&\quad \hat{u}_1^{2^K(i-1)}=z_1^{2^K(i-1)})\\
&= \Pr(\widehat{out}_1 = \vec{\boldsymbol{\alpha}}_{2^K}\boldsymbol{U}(\boldsymbol{1}), \hat{u}_1^{2^K(i-1)}=z_1^{2^K(i-1)})\\
&\quad \times \Pr(\widehat{out}_2 = \vec{\boldsymbol{\alpha}}_{2^K}\boldsymbol{U}(\boldsymbol{2}), \hat{u}_1^{2^K(i-1)}=z_1^{2^K(i-1)})\\
&\quad \times \dots \times \Pr(\widehat{out}_{2^K} = \vec{\boldsymbol{\alpha}}_{2^K}\boldsymbol{U}(\boldsymbol{2^t}), \hat{u}_1^{2^K(i-1)}=z_1^{2^K(i-1)})\\
&= a_1(\vec{\boldsymbol{\alpha}}_{2^K}\boldsymbol{U}(\boldsymbol{1}))a_2(\vec{\boldsymbol{\alpha}}_{2^K}\boldsymbol{U}(\boldsymbol{2}))\dots a_{2^{K-1}}(\vec{\boldsymbol{\alpha}}_{2^K}\boldsymbol{U}(\boldsymbol{2^{K-1}}))\\
&\quad \times b_1(\vec{\boldsymbol{\alpha}}_{2^K}\boldsymbol{U}(\boldsymbol{2^{K-1}+1}))\dots b_{2^{K-1}}(\vec{\boldsymbol{\alpha}}_{2^K}\boldsymbol{U}(\boldsymbol{2^K}))
\end{aligned} \tag{18}
$$

where $\vec{\boldsymbol{\alpha}}_{2^K} \triangleq (\alpha_1, \alpha_2, \dots \alpha_{2^K})$ is a vector consisting of $2^K$ binary 0 or 1.

According to (18), since $P(\alpha_1\alpha_2...\alpha_{2^K})$ is the joint probability of $\hat{u}_{2^K(i-1)+1}=\alpha_1$, $\hat{u}_{2^K(i-1)+2}=\alpha_2$, ..., $\hat{u}_{2^K i}=\alpha_{2^K}$ and $\hat{u}_1^{2^K(i-1)}=z_1^{2^K(i-1)}$, it is just the metric of length-$2^K i$ path $(z_1^{4i-4}\alpha_1\alpha_2...\alpha_{2^K})$. Therefore, with $a_1(0)$, $a_1(1)$,..., $a_{2^{K-1}}(0)$, $a_{2^{K-1}}(1)$, $b_1(0)$, $b_1(1)$,..., $b_{2^{K-1}}(0)$ and $b_{2^{K-1}}(1)$ output from stage-($m$-$K$) and equations (18), MCU can directly output the actual metrics of $2^{2^K}$ length-$2^K i$ paths.

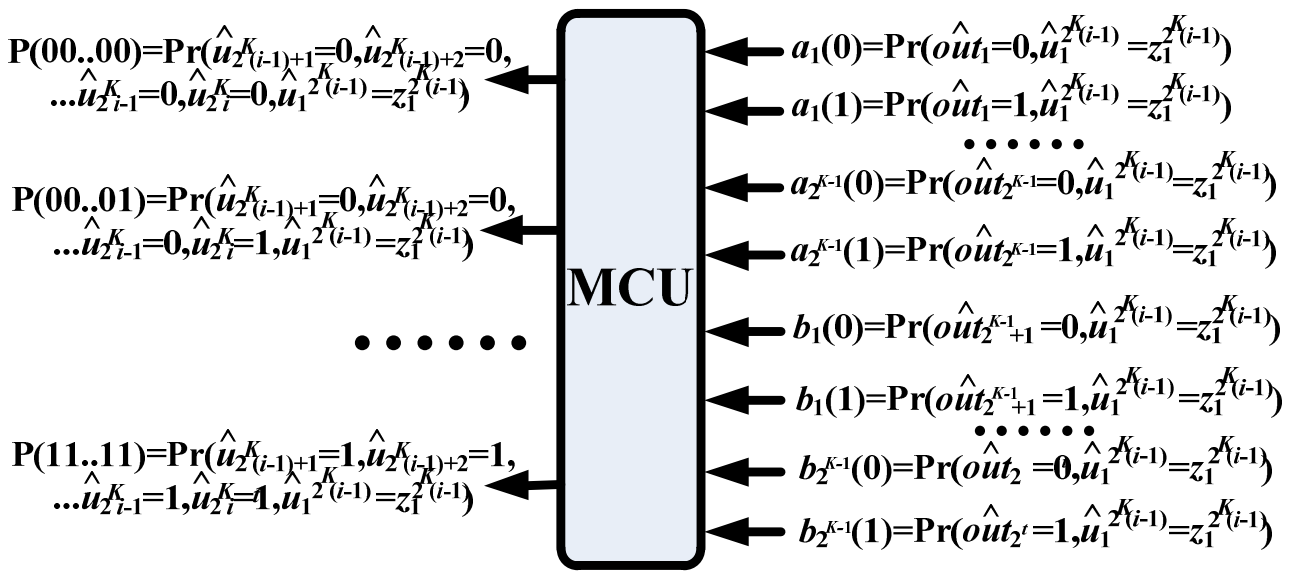

Fig. 13. Block diagram of MCU for $2^K$b-rSCL decoder.

*Zero-Forcing Unit (ZFU)*

Similar to the 2-bit-decision case, the function of ZFU in $2^K$-bit-decision scenario is also to force the metric of unqualified length-$2^K$ paths to 0. Therefore, we can derive the function of ZFU for $2^K$b-rSCL decoder as follows:

Assign $M(\alpha_1\alpha_2...\alpha_{2^K})$ = $P(\alpha_1\alpha_2...\alpha_{2^K})$ for path $(z_1^{4i-4}\alpha_1\alpha_2...\alpha_{2^K})$ with $\alpha_1$, $\alpha_2$,..., $\alpha_{2^K}\in\{0,1\}$;

If $\hat{u}_{2^K(i-1)+1}$ is frozen, then reassign all $M(1\alpha_2\alpha_3...\alpha_{2^K})=0$;

If $\hat{u}_{2^K(i-1)+2}$ is frozen, then reassign all $M(\alpha_1 1\alpha_3...\alpha_{2^K})=0$;

......

If $\hat{u}_{2^K i}$ is frozen, then reassign all $M(\alpha_1\alpha_2...\alpha_{2^K-1}1)=0$ .(19)

Based on MCU in (18) and ZFU in (19), we can develop a general $2^K$b-rSCL decoding algorithm as shown in Scheme-B. Fig. 14 shows the decoding procedure of $2^K$b-rSCL algorithm with list size $L$. It can be seen that during the decoding procedure $2^{2^K}L$ metrics of candidate paths are compared each time, and the $L$ paths with larger $M(\alpha_1\alpha_2...\alpha_{2^K})$ metrics are selected as the survival paths. As a result, $2^K$ successive bits can be determined simultaneously.

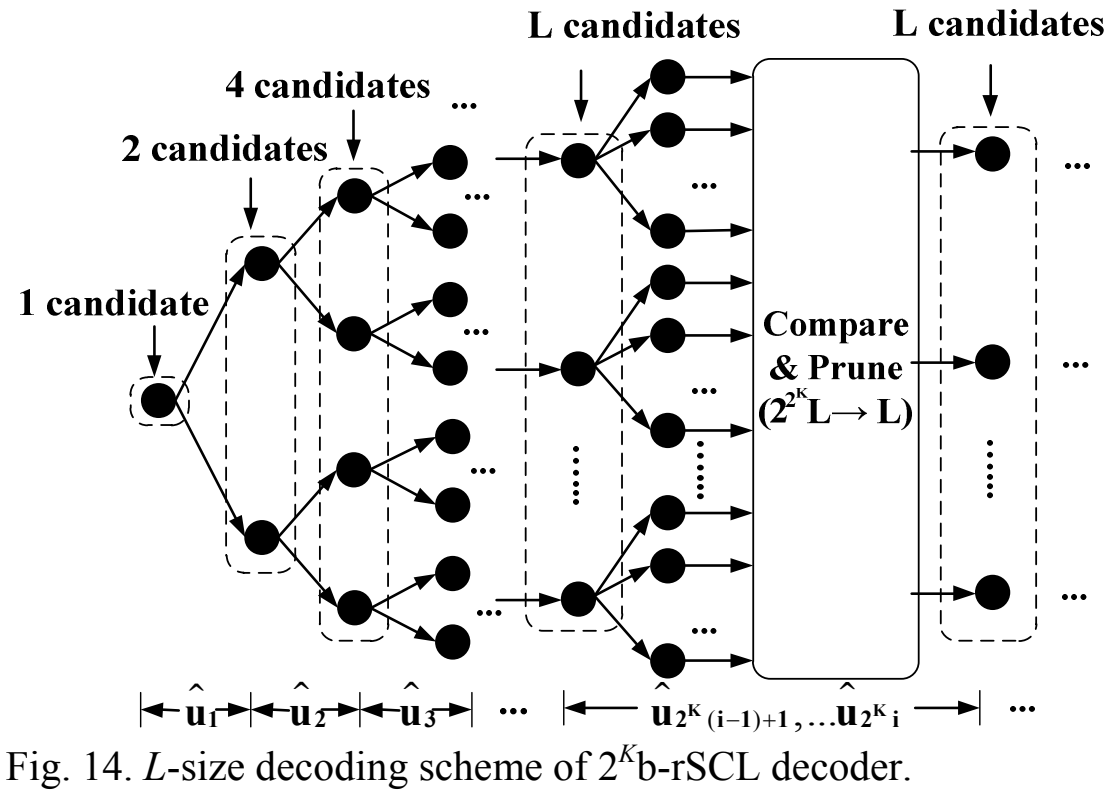

Fig. 14. $L$-size decoding scheme of $2^K$b-rSCL decoder.

**Scheme B: $2^K$-bit SCL decoding ($2^K$b-rSCL) with list size L for (n, k) polar codes**

**Input:** *Likelihoods of each bit in the received codeword*
**For** $i=1$ **to** $n/2^K$
  **For each length-($2^K(i-1)$) survival path** $(\hat{u}_1...\hat{u}_{2^K(i-1)})\triangleq z_1^{2^K(i-1)}$
   **SC decoding:**
    *Activate stage-1 ~ stage-(m-t) of SC component decoder*
    *stage-(m-K) outputs* $a_1(0),a_1(1)...,a_{2^{K-1}}(0),a_{2^{K-1}}(1),b_1(0),b_1(1)...,b_{2^{K-1}}(0),b_{2^{K-1}}(1)$
   **Path Expansion:**
    *Expand survival path* $z_1^{2^K(i-1)}$ *to* $2^{2^K}$ *candidate paths* $(z_1^{2^K(i-1)}\hat{u}_{2^K(i-1)+1}...\hat{u}_{2^K i})$:
    *1 length-(*$2^K(i-1)$*) path* $\Rightarrow 2^{2^K}$ *length-(*$2^K i$*) paths*
   **Metric Computation:**
    *Calculate actual path metrics* $P(\alpha_1\alpha_2...\alpha_{2^K})$ *for* $2^{2^K}$ *length-(*$2^K i$*) paths*:
    $P(\alpha_1\alpha_2...\alpha_{2^K})=a_1(\vec{\alpha}_{2^K}\mathrm{U}(1))a_2(\vec{\alpha}_{2^K}\mathrm{U}(2))...a_{2^{K-1}}(\vec{\alpha}_{2^K}\mathrm{U}(2^{K-1}))$
    $\times b_1(\vec{\alpha}_{2^K}\mathrm{U}(2^{K-1}+1))...b_{2^{K-1}}(\vec{\alpha}_{2^K}\mathrm{U}(2^K))$ *for path* $(z_1^{2^K(i-1)}\alpha_1\alpha_2...\alpha_{2^K})$
   **Forcing Zero:**
    *Calculate the new path metrics* $M(\alpha_1\alpha_2...\alpha_{2^K})$ *with forcing-zero operation*:
    $M(\alpha_1\alpha_2...\alpha_{2^K})=P(\alpha_1\alpha_2...\alpha_{2^K})$ *for path* $(z_1^{2^K(i-1)}\alpha_1\alpha_2...\alpha_{2^K})$ *with* $\alpha_1,...,\alpha_{2^K}\in\{0,1\}$
    $\hat{u}_{2^K(i-1)+1}$ *is frozen* $\Rightarrow$ *all* $M(1\alpha_2\alpha_3...\alpha_{2^K})=0$;
    ......
    $\hat{u}_{2^K i}$ *is frozen* $\Rightarrow$ *all* $M(\alpha_1\alpha_2...\alpha_{2^K-1}1)=0$.
  **End for**
  **Compare and Prune:**
   *Compare metrics* $M(\alpha_1\alpha_2...\alpha_{2^K})$ *of all the* $2^{2^K}L$ *length-(*$2^K i$*) candidate paths*
   *Select L paths with the L largest metrics as the new survival paths*
**End for**
**Output:** *Choose the length-n survival path with the largest metric*

Table III lists the latency of $2^K$b-rSCL decoder with different values of $K$ for ($n$, $k$) polar codes. From this table it can be seen that 2b-rSCL decoder in subsection III-B can be viewed as the specific case of $2^K$b-rSCL with $K$=1. For a general $2^K$b-rSCL decoder, its latency is $n/2^{K-2}$-2 clock cycles. Therefore, as $K$ increases, the overall latency is reduced. In an extreme case, when $K$ reaches $m=\log_2 n$, the $2^K$b-rSCL decoder becomes a maximum likelihood (ML) decoder with latency as small as only 2 cycles.

TABLE III. DECODING LATENCY OF $2^K$B-RSCL DECODER WITH DIFFERENT K

| $K$ | Decoding latency (clock cycles) | Note |
|---|---|---|
| $K$=0 | 3$n$-2 | Original SCL |
| $K$=1 | 2$n$-2 | 2b-rSCL |
| $K$=2 | $n$-2 | 4b-rSCL |
| $K$=3 | $n$/2-2 | 8b-rSCL |
| … | … | … |
| $K$=$K$ | $n/2^{K-2}$-2 | $2^K$b-rSCL (general case) |
| … | … | … |
| $K=m=\log_2 n$ | 2 | Maximum Likelihood (ML) decoder |

Although the increase of $K$ can lead to the reduction of latency, $K$ cannot be set too large for hardware implementation. That is because when $K$ increases, the number of candidate paths, as $2^{2^K}$, increases rapidly. As a result, a large K causes a large amount of path candidates and hence significantly increases the overall complexity of metric computation and path metrics comparison. For example, when $K=m=\log_2 n$ (ML decoder), the number of path candidates is $2^n$. For (1024, 512) polar codes, that means $2^{1024}$ path metrics need to be computed and compared. The implementation of these extensive operations will cause ultra-large silicon area and ultra-long critical path. As a result, for practical implementation $K$ is suggested to be set as no more than 3, which can achieve a good tradeoff between latency reduction and computation overhead.

### *D. Simulation results*

Because the proposed reformulated SCL decoding algorithms only avoid the unnecessary metric computations but do not change the accuracy of metric computation, there is no performance loss for the reformulated SCL algorithms over the original SCL algorithm. This is consistent with the simulation results shown in Fig. 15.

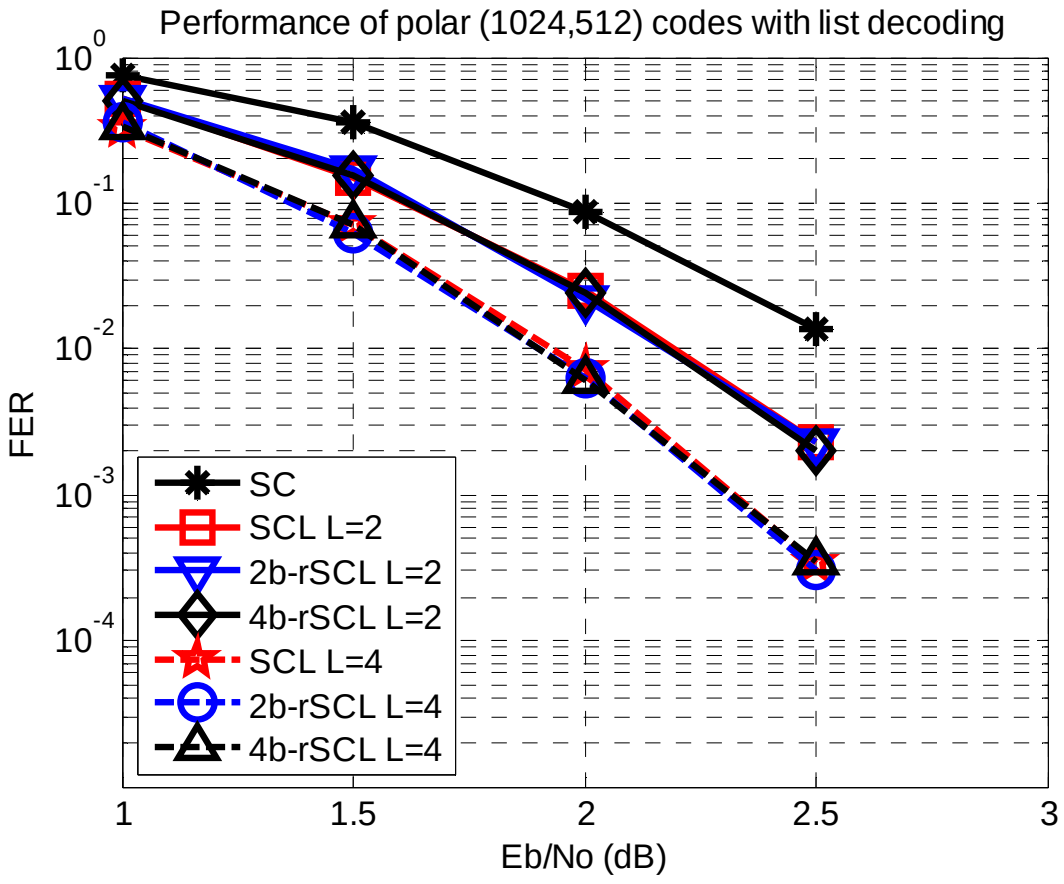

Fig. 15. Performance of $2^K$b-rSCL algorithms for (1024, 512) polar codes.

## IV. The Proposed Reformulated SCL Architecture

In this section, the hardware architectures of the reformulated SCL ($2^K$b-rSCL) decoders are presented. Different values of $K$ correspond to different $2^K$b-rSCL decoders. For simplicity, in this section we focus on $K$=1 and $K$=2 cases, which correspond to the 2b-rSCL decoder and 4b-rSCL decoder. Architectures with values of other $K$ can be developed with a similar way.

As shown in Fig. 11, the difference between SC component decoder of 2b-rSCL or 4b-rSCL decoders and that of original SCL decoder is on the last 1 or 2 stages. Therefore, the other stages (**f**/**g** units) of original SC decoder are still used in the reformulated SCL decoders. As a result, in this section we focus on the architecture design of **f**/**g** units in the SC component decoder, MCU/ZFU in the reformulated stage, and metric sorting block, respectively.

### *A. Processing element for f/g units*

As indicated in Section II, the likelihood-based function of **f** and **g** units are described in (3)(4)(9)(10). However, these equations contain multiplication which is not feasible for hardware implementations. As a result, in order to simplify computation, the log-likelihood-based **f** and **g** units are used in our design. In this case, the likelihood-based (3)(4)(9)(10) are reformulated to the following equations:

$$c(0) = max^*(a(0)+b(0),\ a(1)+b(1)) \quad (20)$$

$$c(1) = max^*(a(0)+b(1),\ a(1)+b(0)) \quad (21)$$

$$d(0) = a(\hat{u}_{sum}) + b(0) \quad (22)$$

$$d(1) = a(1-\hat{u}_{sum}) + b(1) \quad (23)$$

where $max^*(x, y)=max(x, y) + ln(e^{-|x-y|})$ represents the Jacobian logarithm.

Notice that (20) (21) contain logarithmic operation (ln(•)), which needs to be implemented using complex look-up table (LUT) with a long critical path. Fortunately, in [16] it was shown that the logarithmic item can be ignored with negligible performance loss. As a result, (20)(21) can be further simplified as:

$$c(0) = max(a(0) + b(0), a(1) + b(1)) \quad (24)$$

$$c(1) = max(a(0) + b(1), a(1) + b(0)) \quad (25)$$

In general, equations (22)-(25) describe the log-likelihood version of **f** and **g** units. Based on these equations, the basic processing element (PE) of the SC component decoder, which contains an **f** unit and a **g** unit, is developed and is shown in Fig. 16. Here, *C&S* unit represents the combined comparator and 2-to-1 selector. In addition, *ctrl* signal is the control signal that indicates whether the PE functions as an **f** unit or a **g** unit.

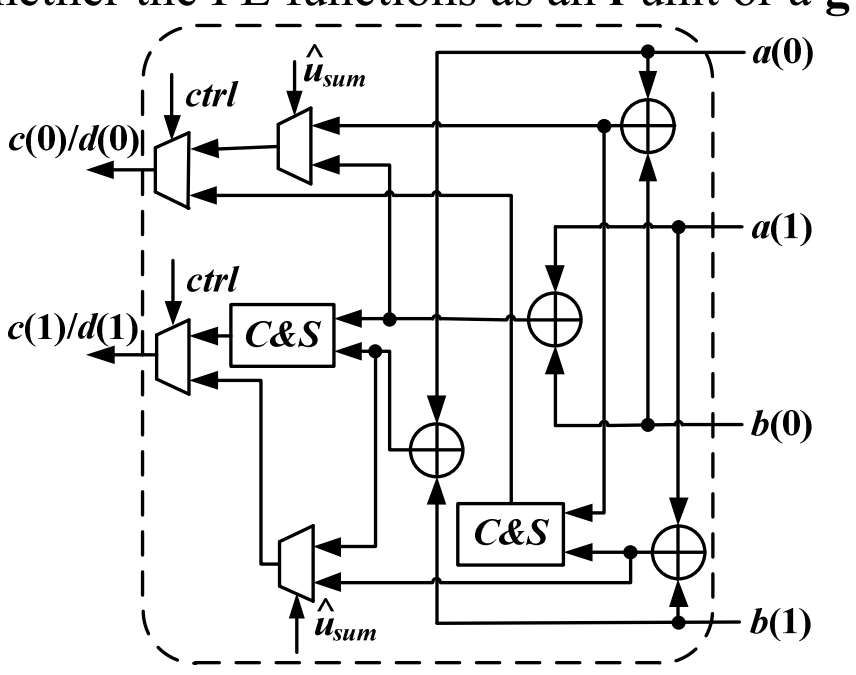

Fig. 16. Architecture of PE for **f** and **g** units in the SC component decoder.

### *B. Metric computation unit (MCU) & Zero-Forcing unit(ZFU)*

As shown in Fig. 11, MCU and ZFU are the two essential parts in $2^K$b-rSCL decoders to help them decide multiple bits. Similar to the case in Section IV-A, the likelihood-based functions of MCU and ZFU need to be transformed to log-likelihood version as well.

For $K$=1 case that corresponds to 2b-rSCL decoding algorithm, its likelihood-based functions of MCU and ZFU have been described in Scheme-A (line10~line18). For the transformation for MCU , according to the transformation principle in Section IV-A, $P(pq)=a(p)b(q)$ in the line-12~line13 of Scheme-A is transformed to $a(p)+b(q)$. In addition, since ln0 is negative infinite, $M(pq)$=0 (line-17~ line-18 in Scheme-A), as the likelihood-based function of ZFU, is reformulated to $M(pq)$=$-Inf$ and where $-Inf$ represents negative infinite. As a result, the hardware architecture of MCU and ZFU for 2b-rSCL decoder is developed as shown in Fig. 17(a). Here the *ctrl1* and *ctrl2* in Fig. 17(a) are the two control signals that indicate whether $\hat{u}_{2i-1}$ and $\hat{u}_{2i}$ are information bits or not.

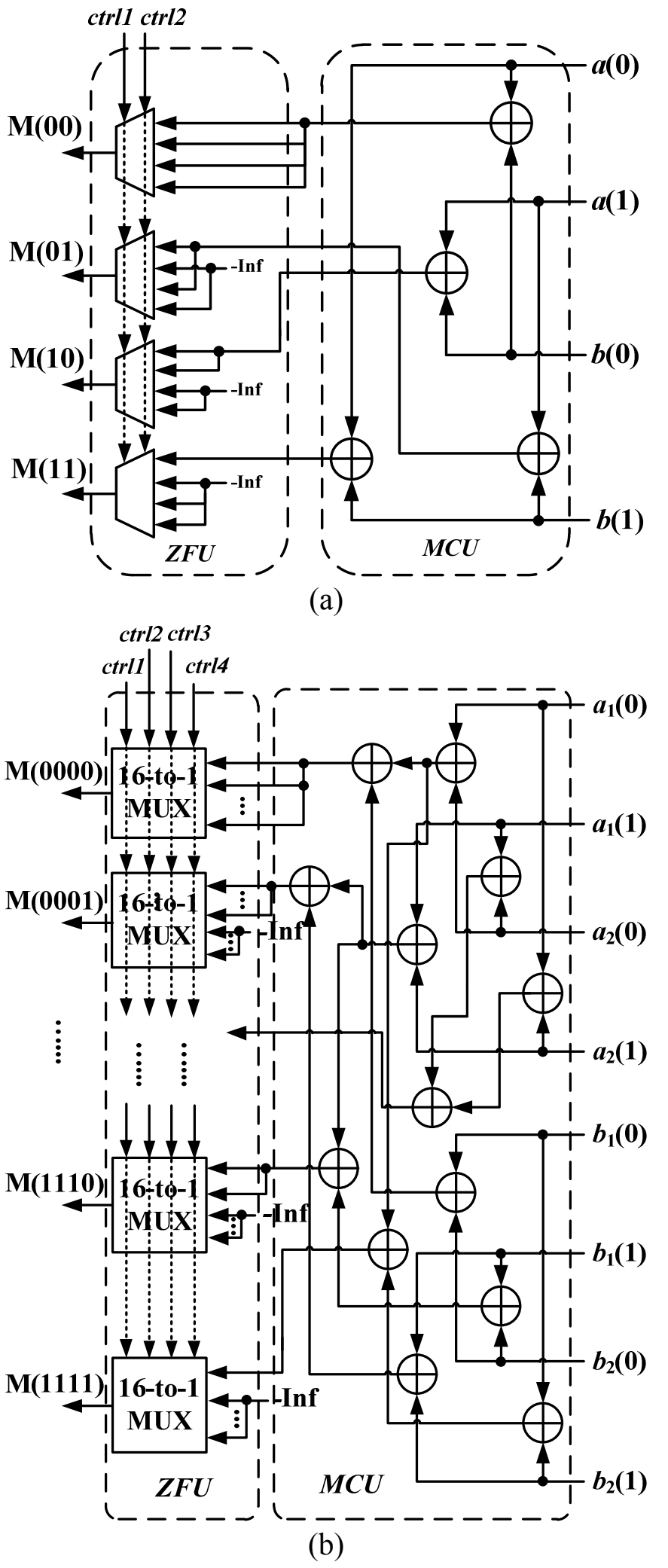

Fig. 17. Architecture of MCU+ZFU for (a) 2b-rSCL (b) 4b-rSCL decoders.

For $K$=2 case that corresponds to 4b-rSCL decoding algorithm, its likelihood-based function of MCU and ZFU can be derived from Scheme-B (line10~line19). For the function of MCU (line12~line13), in $K$=2 case it is $P(\alpha_1\alpha_2\alpha_3\alpha_4)=a_1(\alpha_1 \oplus \alpha_2 \oplus \alpha_3 \oplus \alpha_4)a_2(\alpha_2 \oplus \alpha_4)b_1(\alpha_3 \oplus \alpha_4)b_2(\alpha_4)$. Then, with the likelihood-to-log-likelihood transformation, it is reformulated as $P(\alpha_1\alpha_2\alpha_3\alpha_4)=a_1(\alpha_1 \oplus \alpha_2 \oplus \alpha_3 \oplus \alpha_4)+a_2(\alpha_2 \oplus \alpha_4)+b_1(\alpha_3 \oplus \alpha_4)+b_2(\alpha_4)$. For the function of ZFU (line16~line19), in $K$=2 case it is $M(\alpha_1\alpha_2\alpha_3\alpha_4)=0$. Therefore, its log-likelihood version is $M(\alpha_1\alpha_2\alpha_3\alpha_4)=-Inf$. As a result, the architecture of MCU and ZFU for 4b-rSCL decoders are developed as shown in Fig. 17 (b). Here the *ctrl1, ctrl2, ctrl3* and *ctrl4* in Fig. 17(b) are the four control signals that indicate whether $\hat{u}_{4i-3}$, $\hat{u}_{4i-2}$, $\hat{u}_{4i-1}$ and $\hat{u}_{4i}$ are information bits or not.

## C. Metric Sorting block

After MCU and ZFU generate the metrics for different paths, a sorting block is needed to compare those 2$L$ metrics and select the $L$ paths with larger metrics. In the proposed designs, we use the bitonic sorting algorithm [19] to find out the $L$ larger metrics. Fig. 18 illustrates an example architecture of the proposed 8-input 4-output metric sorting block. It contains a 4x4 increasing order bitonic sorter and a 4x4 decreasing order bitonic sorter. Each bitonic sorter is constructed by 2x2 increasing order sorters (*IOS*) and 2x2 decreasing order sorters (*DOS*). With the help of the two 4x4 bitonic sorters, $in_1$~$in_4$ are sorted as an array with increasing order ($i_1 \le i_2 \le i_3 \le i_4$) while $in_5$~$in_8$ are sorted as an array with decreasing order ($d_1 \ge d_2 \ge d_3 \ge d_4$). Then, these two pre-sorted arrays are sent to a stage of 4 *C&S* units. At the output end of these C&S units, the 4 larger elements among $in_1$~$in_8$ are found as $out_j=max(i_j, d_j)$, where $j$=1, 2, 3 and 4. For the details of bitonic sorter, the reader is referred to [19].

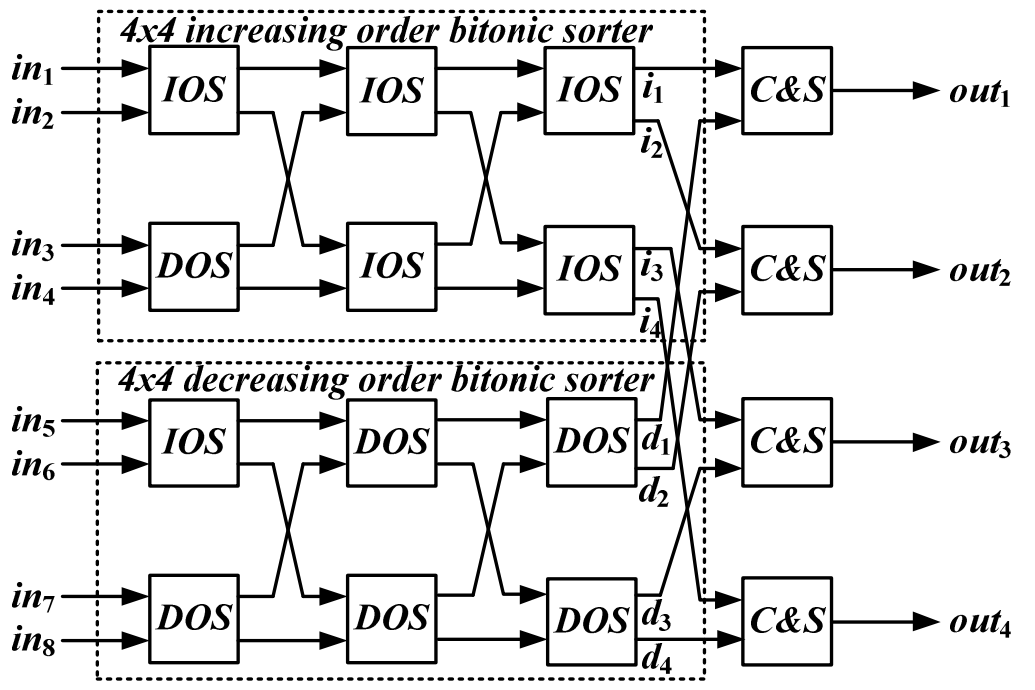

Fig. 18. Architecture of 8-input 4-output metric sorting block.

As indicated in [19], the critical path delay of a $2^s$x$2^s$ bitonic sorter is 1+2+…+$s$=$s$($s$+1)/2 $T_{C\&S}$, where $T_{C\&S}$ is the critical path delay of C&S unit. Therefore, a general $2^s$-input $2^{s-1}$-output metric sorting block consisting of two $2^{s-1}$x$2^{s-1}$ bitonic sorters and a stage of C&S units has an overall critical path delay of 1+2+...$s$-1+1=1+($s$-1)$s$/2 $T_{C\&S}$.

Notice that the metric sorting block is a $2^s$-input $2^{s-1}$-output ($2^s$x$2^{s-1}$) sorter that can find the $2^{s-1}$ largest elements among the $2^s$ inputs. Since for the proposed $L$-size $2^K$b-rSCL decoder, it only needs to find the $L$ largest metrics among $2^{2^K} L$ candidates, hence the $2^s$x$2^{s-1}$ sorter is enough for this sorting task and we do not need the full-size sorting ($2^s$x$2^s$) function.

## D. Data path balancing

As discussed in Section IV-C, the critical path delay of $2^s$-input $2^{s-1}$-output metric sorting block is 1+($s$-1)$s$/2 $T_{C\&S}$. This is much larger than the critical path delay of PE or MCU/ZFU. For example, for a 4b-rSCL decoder with $L$=2, $s$=log$_2$(16*2)=5. Then the critical path delay of metric sorting block is 11$T_{C\&S}$, while the critical path delays of PE and MCU/ZFU are less than 3$T_{C\&S}$. Because the clock speed is upper-bounded by the critical path delay, the throughput of reformulated SCL decoder is limited by the long critical path of metric sorting block.

Considering the unbalanced data path between metric sorting block and other parts of reformulated SCL decoder, we propose to re-pipeline those data paths to reduce critical path delay. Fig. 19(a) shows the original pipelining of 4b-rSCL decoder. Here the register arrays for pipelining are inserted between different blocks of SCL decoder. As a result, due to the unbalanced data path between different blocks, the clock cycles for processing PEs and MCU/ZFU are not fully utilized (see Fig. 19(b)). Fig. 19(c) shows the proposed re-pipelining strategy to the same 4b-rSCL decoder. It can be seen that the original registers between stage-($m$-2), MCU/ZFU and metric sorting block are moved into metric sorting block. Fig. 19(d) shows the corresponding timing chart after re-pipelining. It can be seen that the data path in each clock cycle is balanced. More importantly, since metric sorting block is deeply pipelined, the overall critical path delay is reduced significantly. Notice that in Fig. 19(c) the metric sorting block is 2-stage pipelined. If deeper pipelining is needed, we need to move the registers between other stages of PE into metric sorting block. For example, in order to perform 3-stage pipeline to metric sorting block, we need to move the registers between stage-($m$-3) and stage-($m$-2) in Fig. 19(c) into metric sorting block as well.

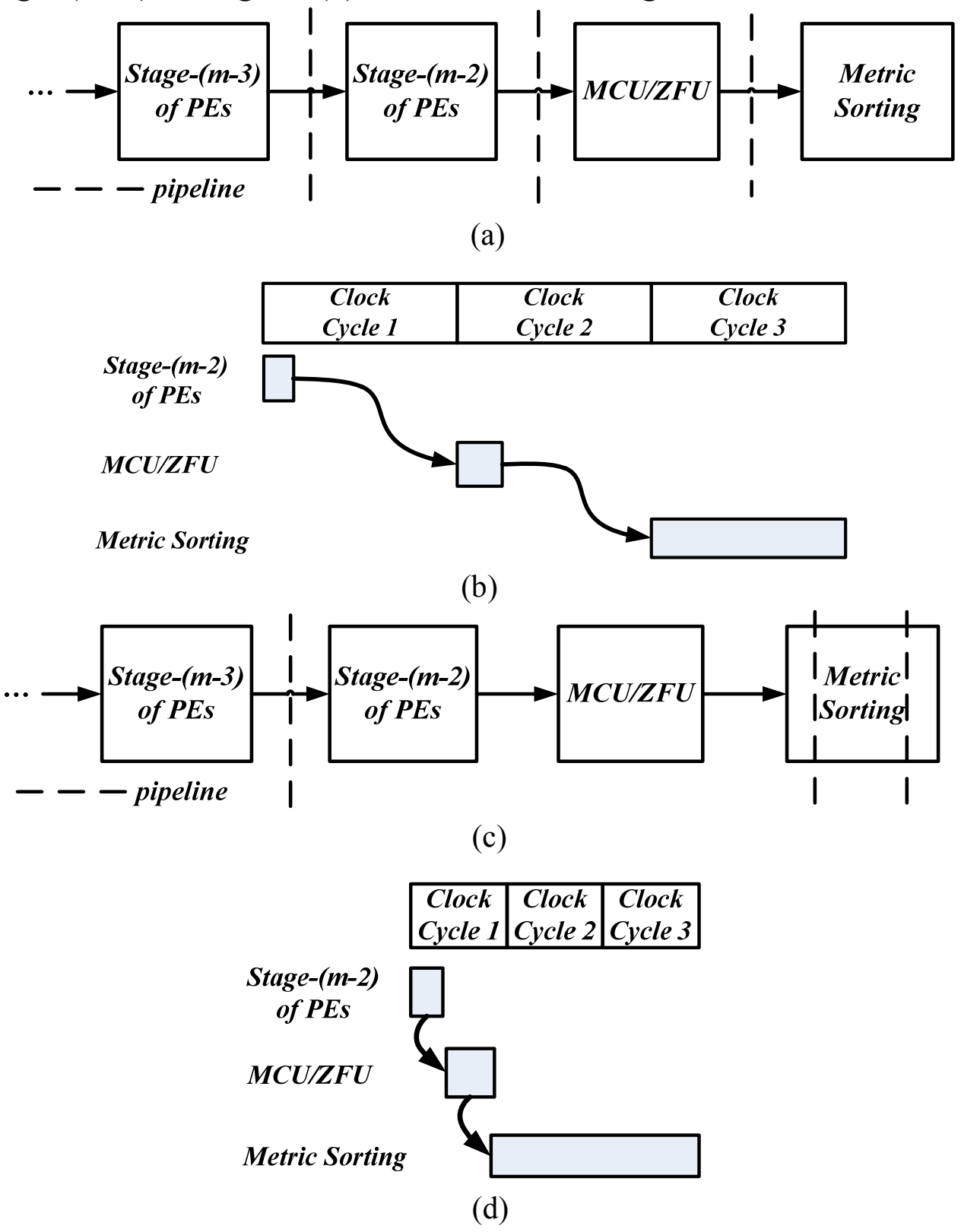

Fig. 19. (a) Original pipelining for 4b-rSCL decoder. (b) Original timing chart. (c) Re-pipelining for 4b-rSCL decoder. (d) Timing chart with balanced data path.

The proposed data path balancing strategy is very useful for high-speed polar list decoder design. For practical use of polar codes, in order to achieve comparable error-correcting performance with LDPC or Turbo codes with the similar codelength, a large list size $L$ is required. For example, [4] reported that the 2048-length polar codes can achieve beyond LDPC performance under the condition of $L$=32. In that case, for the conventional SCL decoder, the $s$ for sorting block is $\log_2(2*32)=6$. As a result, even the proposed metric sorting block is used, the critical path delay is still very large ($(1+(s-1)s/2\ T_{C\&S}=16T_{C\&S}$), which impedes the application of polar codes in high-speed systems. Notice that this phenomenon becomes even more severe for $2^K$b-rSCL decoder. For example, for 4b-rSCL decoder with $L$=32, the number of path metric candidates is 32*16=512, which corresponds to $s=\log_2 512=9$. As a result, the critical path delay of metric sorting block increases to $1+(s-1)s/2\ T_{C\&S}=37T_{C\&S}$. However, if we apply the proposed data path balancing technique to this case, the critical path delay can be significantly reduced. For example, in the case of 2048-length polar codes with $L$=32, with the balance of the data path of metric sorting block, MCU/ZFU block and all the stages of PE (stage-1~stage-9), the critical path delay of 4b-rSCL decoder after data path balancing is less than $(37+3+3*9)/11\approx 6.1T_{C\&S}$. This new critical path delay is 4 times less than the case without use of data path balancing, and it is even 1.5 times less than that of the original SCL decoder. As a result, the use of the proposed data path balancing strategy guarantees the high-speed design of polar list decoder.

## *E. Quantization scheme*

Similar to the case of SCL decoders, the architecture of $2^K$b-rSCL decoders contain multiple stages of PE. As a result, in order to avoid saturation problem that is pointed out in [16], the quantization schemes for different stages of PE are different. If we assume the log-likelihood (LL) information from channel is quantized as $Q_{ch}$ bits, then for the stage-$i$ of $2^K$b-rSCL decoder, the corresponding bit-width is $Q_{ch}+i$. In addition, for the MCU/ZFU and metric sorting blocks, they are quantized with $Q_{ch}+m$ bits. Notice that because the LL information in different stages has different bit-widths, the corresponding memories that store the LL information have different bit-widths as well.

## *F. Memory requirement*

Besides the aforementioned blocks, a large portion of the $2^K$b-rSCL decoders is the memory banks. Similar to SCL decoders [16], multi-bit-width memory banks in the proposed design store the LL information from the channel as well as the LL information processed by each stage. As discussed in the Section IV-E, the quantization scheme for LL information is non-uniform and varies depending on the corresponding stages, therefore the memory banks for different stages have different bit-widths. In addition, 1-bit-width memory banks are needed to store the updated survival paths and partial sum bits $\hat{u}_{sum}$.

Notice that compared to [16], the memory requirement of the proposed $2^K$b-rSCL decoder is larger. This is because the number of path metric candidates increases in the proposed decoders. As a result, more memories are required for storing the calculated metrics from MCU/ZFU block. For example, with $L$=32 and $K$=2, 32*16=512 LL messages for metrics needs to be stored, while SCL decoder only needs to store 64 LL message for metrics. Consider these metrics are always quantized to more than 10 bits, the extra memory requirement of $2^K$b-rSCL decoder causes inevitable area overhead, especially in the case of large $L$ or $K$.

## *G. Overall architecture*

Based on the aforementioned PE, ZFU&MCU and metric sorting block, the overall architecture of an $L$-size reformulated SCL decoder can be developed as illustrated in Fig. 20. Besides the above presented blocks, the decoder needs LL memory bank to store and update the log-likelihood information that are processed by $L$ SC component decoders. In addition, survival path bank is also needed to store and update the $L$ survival paths during the list decoding procedure. Besides that, the reformulated SCL decoder needs a polar-encoder-like partial sum generator (PSG) to compute $\hat{u}_{sum}$ for corresponding SC component decoder. The architecture of PSG is similar to the polar encoder shown in Fig. 1.

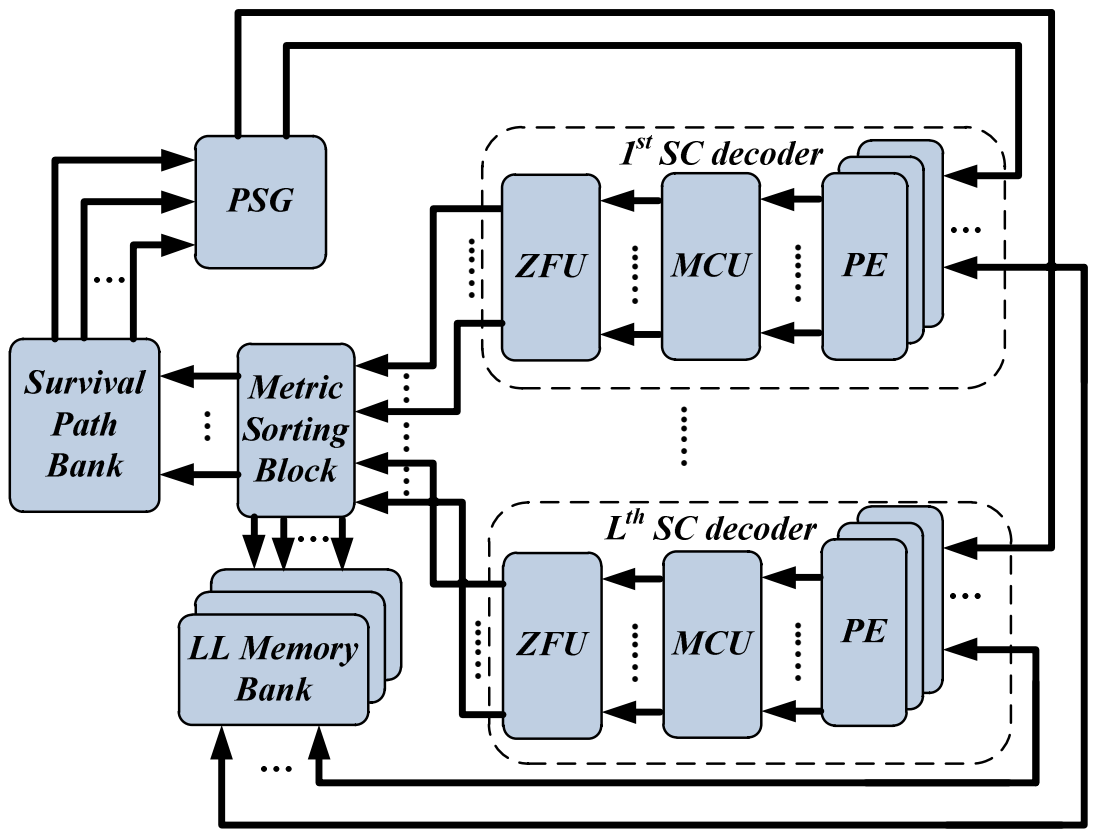

Fig. 20. The overall architecture of reformulated SCL decoders.

## V. Hardware Analysis and Comparison

In this section, the hardware performance characteristics of the proposed reformulated SCL decoding architectures are analyzed. Table IV shows the hardware performance of different SCL decoders with list size $L$=2 and 4 for polar (1024, 512) code. Here the designs of 2b-rSCL decoder and 4b-rSCL decoder are synthesized by Synopsys Design Compiler with ST CMOS 65nm library. Notice that in the proposed designs 3-bit quantization scheme is used for the LL information output from channel, which is the same as in [16]. Based on the quantization scheme described in Section IV-E, the bit width of stage-$i$ is 3+$i$. For the MCU/ZFU block and metric sorting block, they are quantized to 3+$m$=13 bits.

From Table IV it can be seen that, compared with prior LL-based SC list decoder design [16], the proposed 2b-rSCL decoder and 4b-rSCL decoder can achieve 21.0% and 60.5% reduction in latency, respectively. Notice these reductions are less than the analysis in Table III. This is because the latency listed in Table IV is calculated based on the equation (12) in [16], where code rate $R=k/n$ is considered, while the analysis in Table III discuss the general case without the specific discussion on different code rate or distribution of frozen bit positions. In general, as the code rate increases, the proposed reformulated SCL decoders can save more clock cycles than the original one in [16]. For example, for an $R$=1 polar code, 2b-rSCL decoder and 4b-rSCL decoder can achieve 33% and 66% less latency than the original SCL decoder, respectively.

With the use of data path balancing technique in Section IV-D, the proposed 2b-rSCL and 4b-rSCL designs can achieve high clock frequency. Therefore, as seen in Table IV, the coded throughputs of 2b-rSCL decoder and 4b-rSCL decoder with $L$=2 are 1.66 times and 3.45 times of that of original SCL decoder, respectively. In addition, when $L$=4, the coded throughputs of 2b-rSCL decoder and 4b-rSCL decoder are 2.11 times and 3.23 times of that of original SCL decoder, respectively. Besides, the hardware efficiency of our designs, which is defined as the ratio of throughput to area, increases as well. When $L$=2, the hardware efficiencies of 2b-rSCL and 4b-rSCL decoders are 1.36 times and 2.08 times of that of original SCL decoder; when $L$=4, the hardware efficiencies of 2b-rSCL and 4b-rSCL decoders are 1.87 times and 2.66 times of that of original SCL decoder.

Recently, log-likelihood-ratio (LLR)-based SCL decoder was proposed in [17], which requires much less bit-width than LL-based decoder. As a result, the overall area and critical path delay can be significantly reduced. Due to the generality of LLR-based scheme in [17], it can be also applied to our proposed $2^K$b-rSCL decoders. In that case, the hardware complexity and crucial path of our designs can be further reduced while retaining the same short latency.

## VI. Conclusion

In this paper we have presented reformulated SC list decoding algorithms. These reformulated algorithms can reduce the latency significantly without any performance loss. Then, based on the proposed algorithm, we develop corresponding latency-reducing hardware architectures for SCL decoders. Hardware analysis shows that the proposed 2b-rSCL decoder and 4b-rSCL decoder can achieve significant improvement in throughput and hardware efficiency.

TABLE IV. HARDWARE PERFORMANCE OF DIFFERENT (N=1024, K=512) SC LIST DECODERS WITH LIST SIZE L=2,4

| Hardware | SCL [16] | 2b-rSCL | 4b-rSCL | SCL [16] | 2b-rSCL | 4b-rSCL |
|---|---|---|---|---|---|---|
| **List size** | 2 | 2 | 2 | 4 | 4 | 4 |
| **CMOS technology (nm)** | 90 | 65 | 65 | 90 | 65 | 65 |
| **Area($mm^2$) (scaled to 65nm)** | 0.8 | 0.97 | 1.06 | 1.76 | 1.98 | 2.14 |
| **Clock frequency (MHz)** | 459 | 600 | 500 | 314 | 525 | 400 |
| **Latency (clock cycles)** | 2592* | 2046 | 1022 | 2592* | 2046 | 1022 |
| **Coded Throughput (Mbps)** | 181 | 300 | 501 | 124 | 262 | 401 |
| **Hardware efficiency ($Mbps/mm^2$)†** | 226.2 | 309.2 | 472.6 | 70.4 | 132.3 | 187.3 |
| **Power Consumption (mW)** | N/A | 321 | 395 | N/A | 734 | 718 |

* Decoding latency of [16] is calculated based on the equation (12) in [16].

† Hardware Efficiency is defined as Throughput/Area.